\documentclass[
 amssymb, amsmath,
 reprint,%
aps, prl,
superscriptaddress,%
floatfix, showkeys
]{revtex4-2}

\usepackage{graphicx}% Include figure files

\usepackage{pdfpages}
\makeatletter
\AtBeginDocument{\let\LS@rot\@undefined}
\makeatother

\usepackage[T1]{fontenc} % Use modern font encodings
\usepackage[utf8]{inputenc}
\usepackage{lmodern}

\usepackage{subcaption}
\usepackage{multirow}
\usepackage{etoolbox}
\usepackage[separate-uncertainty=true]{siunitx}
\DeclareSIUnit{\kjoule}{\kJ\per\mol}
\DeclareSIUnit{\molecule}{molecule}
\DeclareSIUnit{\kunit}{\cubic\cm\per\molecule\per\second}
\usepackage{upgreek}
\usepackage{chemmacros}

\chemsetup{
 formula = mhchem,
greek = chemgreek,
 modules = thermodynamics,
 modules = redox,
 modules = reactions
}

\renewrobustcmd{\bfseries}{\fontseries{b}\selectfont}
\renewrobustcmd{\boldmath}{}

\usepackage{float}
\usepackage{chemformula} % Formula subscripts using \ch{}
\usepackage[version=4]{mhchem}
\usepackage[T1]{fontenc} % Use modern font encodings

\usepackage{booktabs}
\usepackage{multirow}
\usepackage{xr-hyper}
\usepackage[colorlinks=true,linkcolor=blue]{hyperref}
\newcommand{\etal}{\textit{et al.}}
\newcommand{\cm}{$\mathrm{cm^{-1}}$}
\newcommand{\kjoule}{$\mathrm{kJ~mol^{-1}}$}

\newcommand{\dEr}{$\Delta_r E(0~K)$}

\newcommand{\Ra}{\ce{RuO}}
\newcommand{\Rb}{\ce{RuO2}}
\newcommand{\Rc}{\ce{RuO3}}
\newcommand{\Rd}{\ce{RuO4}}

\newcommand{\Ru}{Ruthenium}
\begin{document}

\title{Reactivity of Ru oxides with air radiolysis products investigated by theoretical calculations}

\author{Faoulat Miradji}
\affiliation{Institut de Radioprotection et de Sûreté Nucléaire (IRSN) PSN-RES, Cadarache, St Paul Lez Durance, 13115, France}
\affiliation{Univ. Lille, CNRS, UMR 8522 - PC2A - Physicochimie des Processus de Combustion et de l'Atmosph\`ere, F-59000 Lille, France}
\affiliation{Laboratoire de Recherche Commun IRSN-CNRS-Lille1 Cin\'etique Chimique, Combustion, R\'eactivit\'e ($\mathrm{C^3R}$), Cadarache, Saint Paul Lez Durance,13115, France}
\affiliation{Univ. Lille, CNRS, UMR 8523 - PhLAM - Physique des Lasers Atomes et Mol\'ecules, F-59000 Lille, France}

\author{Sidi M. O. Souvi}
\affiliation{Institut de Radioprotection et de Sûreté Nucléaire (IRSN) PSN-RES, Cadarache, St Paul Lez Durance, 13115, France}
\affiliation{Laboratoire de Recherche Commun IRSN-CNRS-Lille1 Cin\'etique Chimique, Combustion, R\'eactivit\'e ($\mathrm{C^3R}$), Cadarache, Saint Paul Lez Durance,13115, France}

\author{Laurent Cantrel}
\affiliation{Institut de Radioprotection et de Sûreté Nucléaire (IRSN) PSN-RES, Cadarache, St Paul Lez Durance, 13115, France}
\affiliation{Laboratoire de Recherche Commun IRSN-CNRS-Lille1 Cin\'etique Chimique, Combustion, R\'eactivit\'e ($\mathrm{C^3R}$), Cadarache, Saint Paul Lez Durance,13115, France}

\author{Florent Louis}
\affiliation{Univ. Lille, CNRS, UMR 8522 - PC2A - Physicochimie des Processus de Combustion et de l'Atmosph\`ere, F-59000 Lille, France}
\affiliation{Laboratoire de Recherche Commun IRSN-CNRS-Lille1 Cin\'etique Chimique, Combustion, R\'eactivit\'e ($\mathrm{C^3R}$), Cadarache, Saint Paul Lez Durance,13115, France}

\author{Valérie Vallet}
\affiliation{Univ. Lille, CNRS, UMR 8523 - PhLAM - Physique des Lasers Atomes et Mol\'ecules, F-59000 Lille, France}

\date{Received: \today}

\begin{abstract}
Quantitative predictions of the release of volatile radiocontaminants of ruthenium (Ru) in the environment from either nuclear power plants (NPP) or fuel recycling accidents present significant uncertainties while estimated by severe accidents nuclear analysis codes. Observations of Ru from either experimental or modeling works suggest that the main limitations relate to the poor evaluation of the kinetics of gaseous Ru in the form of {\Rc} and {\Rd}. This work presents relativistic correlated quantum chemical calculations performed to determine the possible reactions pathways leading to the formation of gaseous Ru oxides under NPP severe accident conditions, as a result of reactions of {\Rb} gaseous with air radiolysis products, namely nitrous and nitrogen oxides. The geometries of the relevant species were optimized with the TPSSh-5\%HF functional of the density, while the total electronic energies were computed at the CCSD(T) level with extrapolations to the complete basis set CBS limit. The reaction pathways were fully characterized by localizing the transition states and all intermediate structures using the internal coordinate reaction algorithm (IRC). The rate constants were determined over the temperature range \SIrange{250}{2500}{\kelvin}. It is revealed that the less kinetically limiting pathway to form Ru gaseous fraction is the oxidation of Ru by nitrogen oxide, corroborating experimental observations.
\end{abstract}

\maketitle

\section{Introduction}

The prevention of the radiological consequences of ruthenium release in the environment, implying the evaluation of its source term, is crucial for nuclear safety, as ruthenium-containing compounds present severe sanitary issues due to \ce{^{103}Ru} and \ce{^{106}Ru} isotopes, considered as short and medium radio contaminants like \ce{^{131}I} and \ce{^{137}Cs}~\cite{cea03}. Such a release may occur mostly in the event of a nuclear power plant (NPP) severe accident (SA) like the Three Mile Island or the Chernobyl ones~\cite{VanDorsselaere201519,ruthenium-Pontillon-NED2010-240-1867}, but sometimes from nuclear fuel recycling plant as observed lately through the Ural accident provoking a release of ruthenium \ce{^{106}Ru} over European countries~\cite{ru2017a,ru2017b}. Literature review on Ru transport studies reported that Ru volatilization can be influenced by humidity, temperature, and air flow rate in the Reactor Coolant System (RCS)~\cite{ruthenium-Karkela-ANE2014-74-173,ruthenium-Ohnet-RNC2018-316-161} during a severe accident. The interactions with other elements released from the fuel may also impact the chemical composition and possibly the quantity of transported Ru~\cite{DiLemma2015499}. The mechanisms involved in the ruthenium release are still not well characterized and implemented in SA analysis codes~\cite{miraERMSAR}. The volatile form of Ru is ruthenium tetroxide that can exist in oxidizing atmosphere and which is not easy to trap by usual filters~\cite{ruthenium-Nerisson-JRNC2019-321-591} but can be thermally reduced into non-volatile {\Rb} form. Chemical reactivity of gaseous {\Rd} has to be modeled in order to better assess potential releases in the frame of a nuclear accident. 
In a previous theoretical work, we have used relativistic correlated quantum chemical methods to consolidate and extend Ru thermodynamics database~\cite{mira15,mira16}. It was revealed that thermodynamic properties of Ru compounds in SA databases agree with our theoretical calculations. 
Thence, a possible explanation for the observed discrepancies between the simulated and experimental amounts of Ru transported fractions, might be related to kinetic limitations in the formation of gaseous ruthenium molecules~\cite{miradji-thesis}.

Thus, several research programs are currently led to improve knowledge on Ru transport schemes~\cite{start10,start14,ruthenium-Karkela-ANE2014-74-173,vtt-kajan,kajan2016}.
Among them, the Technical Research Centre of Finland (Teknologian tutkimuskeskus, VTT) program studied ruthenium transport in humid atmosphere with air radiolysis products~\cite{ruthenium-Karkela-ANE2014-74-173,vtt-kajan,kajan2016}. The radiation of the atmosphere in the reactor vessel is thought to occur under an NPP SA leading in particular to the formation of air radiolysis products like nitrogen and/or nitrous oxides~\cite{grahm78,gir2002}.  
The results of the VTT program underlined that the major part of Ru released from the crucible was deposited inside the apparatus. The transported Ru fractions were in condensed and gaseous forms. The aerosols were identified as {\Rb} and the gaseous form was identified as~{\Rd}, according to instrumental neutron activation analysis. As it was found that air radiolysis products enhance the fraction of gaseous Ru at the outlet of the experimental set-up, we decided to launch quantum chemical calculations to determine reaction pathways and their kinetic parameters that lead to the formation of the gaseous fraction of ruthenium ({\Rc} and {\Rd} compounds, likely to be formed under experimental conditions~\cite{mira16}), with air radiolysis products \ce{N2O} and \ce{NO2}, according to the following chemical reactions:
\begin{align}
\label{eqru-n2a}
          \ce{RuO2 +  N2O  &-> RuO3 + N2}  \\
\label{eqru-n2b}        
         \ce{RuO3 +  N2O  &-> RuO4 + N2}  \\
\label{eqru-noa}       
         \ce{RuO2 +  NO2  &-> RuO3 + NO} \\
\label{eqru-nob}       
         \ce{RuO3 +  NO2  &-> RuO4 + NO}.
\end{align}
\label{reacNOX}

\section{Computational methodology}

\subsection{Structural properties and energetics} 

The geometries of all stationary points for the relevant chemical reactions were optimized using the TPSSh-5\%HF density functional~\cite{dft-Tao-PRL2003-91-146401}, validated in our previous works to obtain thermodynamic properties of~{\Ru} oxides and oxyhydroxides~\cite{mira15,mira16}.  In our calculations, the N, O, and H atoms are described with augmented correlation consistent polarized triple zeta aug-cc-pVTZ (noted aVTZ) basis sets~\cite{dunning89}. The ruthenium atom is described by the aug-cc-pVTZ-PP basis set of Peterson~{\etal}~\cite{ecp-Peterson-JCP2007-126-124104}, which incorporates a relativistic pseudo-potential that accounts for scalar relativistic effects in the Ru atom. Vibrational frequencies and zero-point energies (ZPE) were determined at the same level of theory as the geometries. When a transition state structure has been located on the potential energy surface, intrinsic reaction coordinates calculations (IRC) have been performed using the algorithm implemented in the Gaussian09 RevC.01 software~\cite{g09} to find the associated molecular complexes on the reactant (MCR) and product (MCP) sides. 

In a second step, to obtain accurate potential energies of all stationary points, single and double coupled cluster with inclusion of a perturbative estimation for triple excitation (CCSD(T))~\cite{prog-Knowles-JCP1993-99-5219,prog-Knowles-JCP2000-112-3106} calculations were performed with the MOLPRO quantum chemistry package Version 2015.1~\cite{MOLPRO-WIREs,prog-molpro}. The full valence (oxygen 2s and 2p and ruthenium 5s and 4d) orbitals of Ru compounds were correlated in the CCSD(T) calculations. The total energies computed with triple-$\zeta$ ($n = 3$), and quadruple-$\zeta$  ($n=4$) basis set qualities were extrapolated to the complete basis set (CBS) limit using a two-fold scheme to extrapolate Hartree-Fock energies~\cite{feller92,feller93} and correlation energies~\cite{helgaker97}, as described in our previous works~\cite{mira15,mira16}. The use of DFT optimized geometries combined with CCSD(T) single point energy calculations was portrayed in a recent work by Fang~{\etal}~\cite{dix17} to study small Ru clusters with water, enhancing the appropriateness of such methodology to evaluate Ru related compounds thermochemistry.

Regarding spin-orbit (SO) coupling, it was found in our previous investigation~\cite{mira15} that SO contributions amounted to \SIlist{16; 10; 0.75}{\kjoule} for Ru, {\Ra}, and {\Rb} compounds, respectively. In this study, we don't  expect SO contributions to exceed that found for the Ru atom, and as the ground state wave functions have low spin contamination, we can safely assume that SO corrections cancel when considering reaction and activation energies. 

\subsection{Rate constants}
The rate constants of the reactions were calculated using the direct transition state approach, as applied in our previous work~\cite{doi:10.1021/acs.jpca.6b00047}. The direct mechanism considers the reaction from the reactants to the products, and the formation of the pre-reactive complex (MCR) is disregarded. The canonical Transition State Theory (TST)~\cite{johnston66,laidler69,weston72,rapp72,nikitin74,smith80,steinfeld89} was applied to calculate the temperature dependence of the rate constant
for the direct mechanism, $k_{direct}$, as follows:
\begin{align}
k_{direct} (T) = & \Gamma (T)\times \frac{k_BT}{h}\times \frac{Q_{\ce{TS}}(T)}{Q_{\ce{A}}(T)Q_{\ce{B}}(T)}\nonumber\\
& \times \exp{\left(-\frac{E_{\ce{TS}} - E_{\ce{A}} - E_{\ce{B}}}{k_BT}\right)}
\label{eq:kdir}
\end{align}
where $\Gamma (T)$ represents the transmission coefficient used for the tunneling correction at temperature $T$, and $k_B$ and $h$ are the Boltzmann and Planck constants, respectively. $Q_{\ce{A}} (T)$, $Q_{\ce{B}}(T)$, and $Q_{\ce{TS}}(T)$ are the total partition functions of A, B, and the TS at the temperature $T$, respectively. $E_{\ce{A}}$, $E_{\ce{B}}$, and $E_{\ce{TS}}$ are the total energies at \SI{0}{\kelvin} including the zero-point energies.

%The indirect mechanism considers the formation of intermediate reactant MCR.
%The equilibrium reaction of formation/dissociation of the pre-reactive complex is fast. The overall rate constant for the indirect mechanism, $k_{indirect}$, can be written as:\\
%\begin{equation}
%k_{indirect} (T)= K_{a,b}(T)\times k_c(T)
%\end{equation}\label{kind}
%where $K_{a,b} = \displaystyle\frac{k_a }{k_b}$ is the equilibrium constant between the isolated reactants and the pre-reactive complex MCR and $k_c$ is the rate constant for the reaction from MCR to TS. $K_{a,b}$ and $k_c$ are calculated using the following relations:\\
%\begin{equation}
%K_{a,b}(T) = \frac{Q_{\ce{MCR}}(T)}{Q_{\ce{A}}(T)Q_{\ce{B}}(T)}\times \exp{\left(-\frac{E_{\ce{MCR}}-E_{\ce{A}} - E_{\ce{B}}}{k_BT}\right)}
%\end{equation}
%\begin{equation}\label{kind1}
%k_c(T) =  \Gamma (T)\times \frac{k_BT}{h}\times \frac{Q_{\ce{TS}}(T)}{Q_{\ce{MCR}}(T)}\times \exp{\left(-\frac{E_{\ce{TS}} - E_{\ce{MCR}}}{k_BT}\right)},
%\end{equation}\label{kind2}
%where $Q_{\ce{MCR}} (T)$ is the total partition function of the pre-reactive complex at the temperature $T$. $E_{\ce{MCR}}$ is the total energy at \SI{0}{\kelvin} including the zero-point energy.

The GPOP program~\cite{gpop} was used to extract information from the Gaussian output files to estimate the Eckart tunneling corrections and to perform the rate constant calculations over the temperature range of interest, \SIrange{250}{2500}{\kelvin}. The structural properties, energetics and  kinetic parameters of selected reactions pathways are discussed in the next subsections.

\section{Results and discussion}

% \subsection{Geometrical parameters of potential energy surfaces}\label{geom-pes}

%The Gibbs energy curves for the reactions~\ref{eqru-n2a},~\ref{eqru-n2b},~\ref{eqru-noa}, and~\ref{eqru-nob} are provided in supplementary material (Fig.~\ref{test2}), along with the optimized Cartesian coordinates, rotational constants, zero-point energy corrections (ZPE), point group, symmetry number, and frequencies ($\omega$) of reactants, products (Table~\ref{test3}) and intermediate species (Tables~\ref{test4},~\ref{test5},~\ref{test6} and~\ref{test7}-\ref{test10} for equation~\ref{eqru-n2a},~\ref{eqru-n2b},~\ref{eqru-noa} and~\ref{eqru-nob}, respectively). The computed $r$(\ce{N-N}) bond length for the \ce{N2} species with the TPSSh-5\%HF functional is \SI{1.097}{\angstrom}, and is very close to the tabulated experimental value in the NIST database~\cite{hub79} (\SI{1.098}{\angstrom}). In the linear \ce{N2O} molecule, this (\ce{N-N}) bond length is found to be equal to \SI{1.130}{\angstrom}, and the (\ce{N-O}) to \SI{1.187}{\angstrom}, in agreement with the literature values of \SIlist{1.128;1.184}{\angstrom}~\cite{herz66}. From Figure~\ref{test2}, we could affirm that all reactions are spontaneous in standard conditions, having all negative-free energies. However, the oxidation reactions involving \ce{N2O} species seem to request less energy than those involving \ce{NO2} species. To get more insight onto these mechanisms, the potential energy surfaces of each reaction and their kinetic parameters are discussed hereafter. 

\subsection{Reaction coordinates with \ce{N2O}}

\begin{figure}
\includegraphics[width=\linewidth, trim={0cm 0cm 0cm 1.3cm },clip]{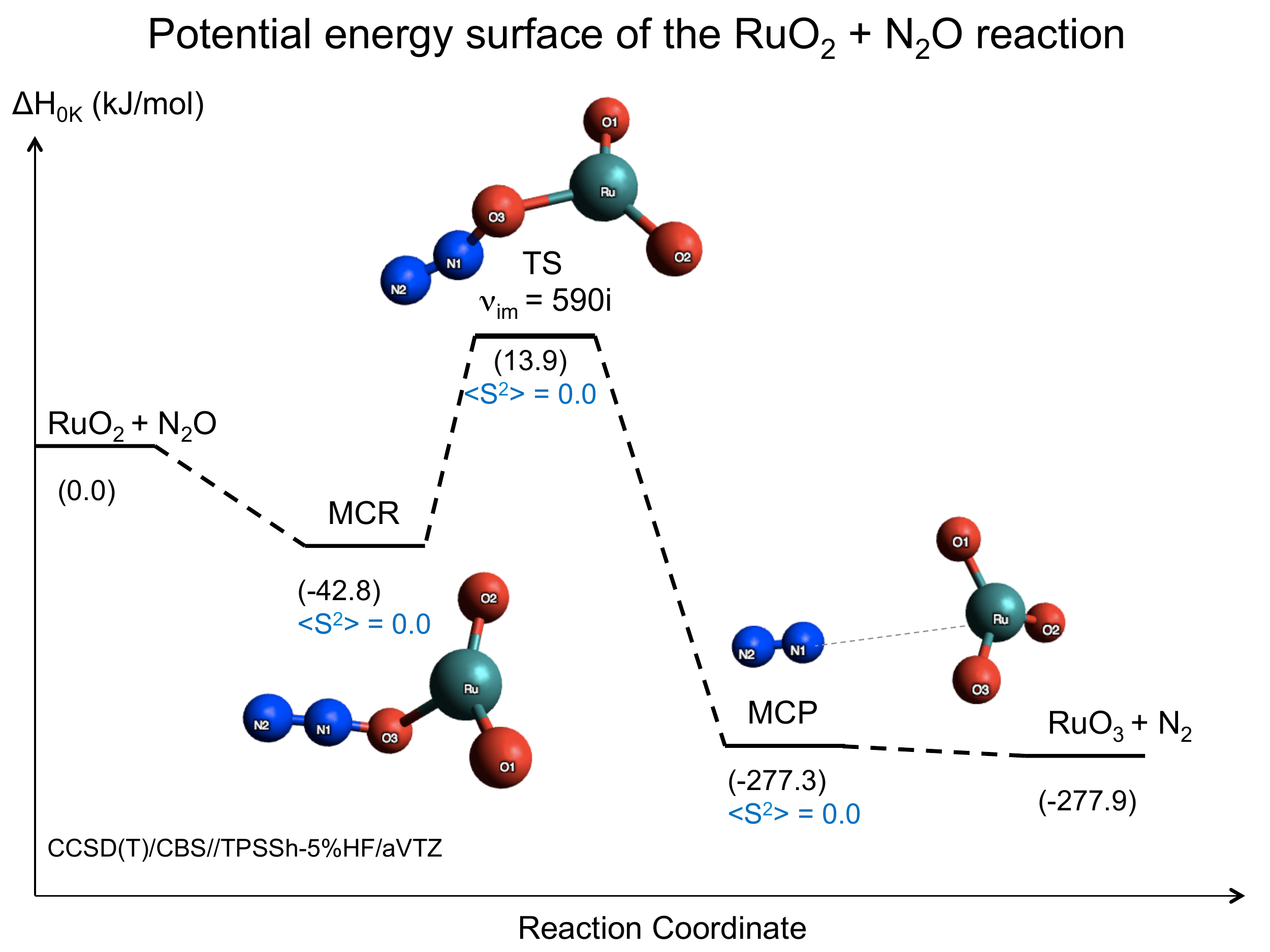}\\ 
\caption{\small Reaction coordinate at \SI{0}{\kelvin} of the \ce{RuO2 + N2O -> RuO3 + N2} reaction, including ZPE, calculated at the CCSD(T)/CBS//TPSSh-5\%HF/aVTZ level of theory, displayed along with schematic representations of the intermediate species involved.}
\label{fig:reacN2or3}
\end{figure}

\begin{figure}
\includegraphics[width=\linewidth, trim={0cm 0cm 0cm 1.3cm },clip]{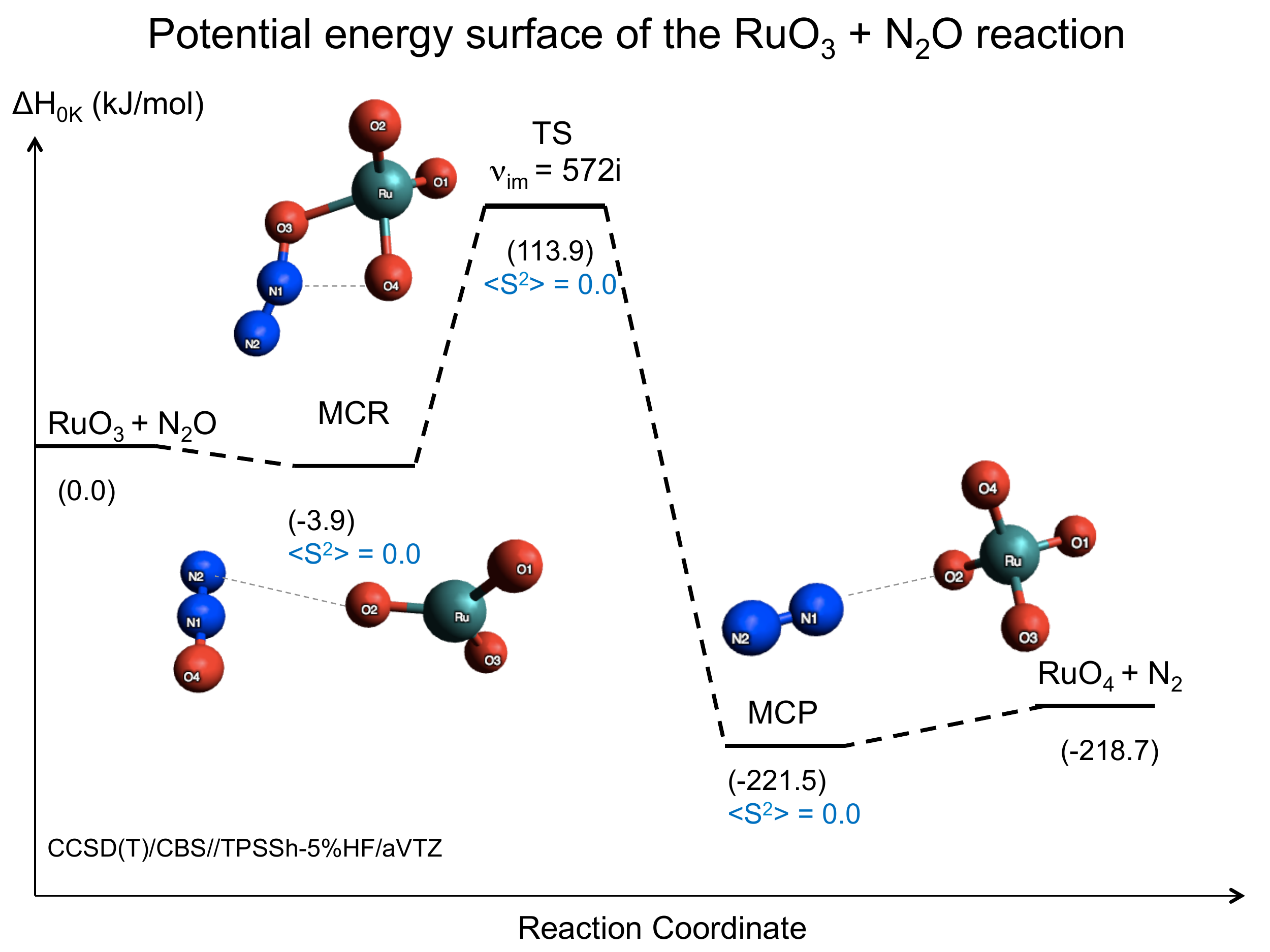}\\ 
\caption{\small Reaction coordinate at \SI{0}{\kelvin} of the \ce{RuO3 + N2O -> RuO4 + N2} reaction, including ZPE, calculated at the CCSD(T)/CBS//TPSSh-5\%HF/aVTZ level of theory, displayed along with schematic representations of the intermediate species involved.}
\label{fig:reacN2or4}
\end{figure}

The reaction coordinates involving oxidation of {\Rb} by \ce{N2O} are displayed in Fig.~\ref{fig:reacN2or3} for the formation of {\Rc} and Fig.~\ref{fig:reacN2or4} for the formation of {\Rd}. The corresponding geometrical 
parameters and ZPE corrections are listed in Tables~\ref{tab:runxo-react2a}, and~\ref{tab:runxo-react2b}, respectively.

\begin{table}
\centering
\caption{ \small Structural parameters (bond lengths $r$ in~\si{\angstrom}), imaginary vibrational frequency (\si{\per\cm}), and ZPE (\kjoule), for the transition state and molecular complexes calculated at the TPSSh-5\%HF/aVTZ  level of theory, involved in the reaction \ce{RuO2 + N2O  -> RuO3 +N2}}
\begin{tabular}{lccc}
\toprule
Parameters &            MCR(\ref{eqru-n2a})            &   TS(\ref{eqru-n2a})          & MCP(\ref{eqru-n2a}) \\
\midrule
$r(\ce{N-N})$               & 1.124                & 1.128  & 1.097\\
$r(\ce{N-O})$              & 1.222                & 1.351  & \\
$r(\ce{Ru-O_{N_2O}})$ &  2.126   & 1.934     & \\
$r(\ce{Ru-O})$                  & 1.680  & 1.676 & 1.685\\
$r(\ce{Ru-N_{N_2}})$ &                 &         & 4.527\\
$\theta(\ce{N-N-O})$  & 173.4  & 148.8  & \\
$\theta(\ce{O-Ru-O_{N_2O}})$ & 109.8  & 113.9 \\
$\theta(\ce{O-Ru-O})$    &    140.5 &  133.9   & 120.0 \\
$\nu_{im}$               &&   590\textbf{i}  & \\
ZPE                        & 43.94  &  49.96 &  36.20 \\
\bottomrule    
\end{tabular}
\label{tab:runxo-react2a}
\end{table}

\begin{table}
\centering
\caption{\small Structural parameters (bond lengths r in~\si{\angstrom}) imaginary vibrational frequency (\si{\per\cm}), and ZPE (\kjoule), for the transition state and molecular complexes calculated at the TPSSh-5\%HF/aVTZ  level of theory, involved in the reaction \ce{RuO3 + N2O -> RuO4 +N2}}
\begin{tabular}{lccc}
\toprule
Parameter &            MCR(\ref{eqru-n2b})             &   TS(\ref{eqru-n2b})           & MCP(\ref{eqru-n2b})  \\
\midrule
$r(\ce{N-N})$               &         1.130       &  1.121 & 1.098 \\
$r(\ce{N-O})$              &        1.187         & 1.262 & \\
$r(\ce{Ru-O_{N_2O}})$ &   4.986  &   2.103  & \\
$r(\ce{Ru-O})$                  &  1. 684 &  1.762-1.688 & 1.684  \\
$r(\ce{Ru-N})$ &                 &         & 4.020\\
$\theta(\ce{N-N-O})$  & 180  &  159.4  & \\
$\theta(\ce{O-Ru-O_{N_2O}})$ & 118.8 & 113.1-118.1 & \\
$\theta(\ce{O-Ru-O})$    &  120.0  & 108.1   &  109.3-109.5 \\
$\nu_{im}$               &   & 572\textbf{i}  & \\
ZPE                        & 50.73  & 49.96  &   48.62 \\
\bottomrule
\end{tabular}
\label{tab:runxo-react2b}
\end{table}


The TS(\ref{eqru-n2a}) structure corresponding to the formation of {\Rc} features a one-step mechanism with the breaking of the \ce{N-O} bond, stretched up to \SI{1.351}{\angstrom} from its equilibrium value in~\ce{N2O}, to form the third Ru-O bond, equal to \SI{1.934}{\angstrom}. This bond length is typical of ionic bonding, as observed between hydroxyl ligands and Ru oxides in our previous work~\cite{mira16}. The value of the {\Rb} bond angle decreases to favor a pyramidal structure leading to the formation of {\Rc}.

In the TS(\ref{eqru-n2b}) structure leading to the formation of {\Rd}, the elongation of \ce{N-O} bond distance is only \SI{0.075}{\angstrom}, much closer to the equilibrium reactant structure than in the previous TS. The bond angles and bond distances shrink from their equilibrium values, to form the tetrahedral {\Rd} structure. This reaction involves the creation of the fourth \ce{Ru-O} bond with a length of \SI{2.101}{\angstrom}, sharing fewer electrons between Ru and O$_{\ce{N2O}}$ atoms compared to its counterpart in the previous TS. A bridge between O$_{\Rc}$ and N$_{\ce{N2O}}$ is also observed, measuring \SI{2.258}{\angstrom}, indicating that the formation of {\Rd} originates from a large orbital mixing between the reactants.

The connections between TS(\ref{eqru-n2b}) and both the \ce{RuO2\bond{...}N2O} reactant complex (MCR(\ref{eqru-n2a})) and \ce{RuO3\bond{...}N2} product complex (MCP(\ref{eqru-n2a})) have been established in both forward and backward directions via IRC calculations at the TPSSh-5\%HF/aVTZ level of theory. The ionic interaction (\SI{2.126}{\angstrom}) between $\ce{O_{N_2O}}$ and the Ru atom stabilizes the {\Rb} and \ce{N2O} reactants to form the MCR(\ref{eqru-n2a}) complex. In reaction~\ref{eqru-n2b}, the structure of the MCR involves a very weak interaction of the Van der Waals type (\SI{4.986}{\angstrom}), leaving the bond angles and bond distances similar to those in {\Rc} and \ce{N2O} reactant species. Such interactions are also observed in MCP systems for both reactions, where the \ce{Ru-N} bond length is equal to \SIlist{4.527;4.020}{\angstrom}, for \ce{RuO3\bond{...}N2} and \ce{RuO4\bond{...}N2} complexes, respectively.
The relative enthalpies at \SI{0}{\kelvin} in the reaction coordinate curve reveal that the reaction between \ce{N2O} and {\Rb} has to overcome a barrier of about \SI{14}{\kjoule} to form the \ce{N2} and {\Rc} products. The MCR(\ref{eqru-n2a}) is stabilized by $\sim$~\SI{43}{\kjoule} with respect to the reactants. The MCP is similar to the products, and lies only \SI{0.6}{\kjoule} below the product energy limit.

For the reaction {\ref{eqru-n2b}} (See Fig.~\ref{fig:reacN2or4}), the transition state is located above reactants with an important vibrationally adiabatic barrier of about 113.9~{\kjoule} by comparison to the one found for reaction 1a (13.9~{\kjoule}).The pro- and post-reactive complexes are similar to the reactants and products, respectively, differing only by $\sim$~\SI{4}~{\kjoule}.

\subsection{Reaction coordinates with \ce{NO2}}

We now turn to presenting the reaction coordinates corresponding to \ce{NO2} oxidation. 
The optimized geometry parameters for NO ($r(\ce{N-O})\sim$\SI{1.154}{\angstrom}) and \ce{NO2} ($r(\ce{N-O})\sim$\SI{1.199}{\angstrom}; $\theta$(\ce{O-N-O})$\sim$\SI{134.2}{\degree}) are in good agreement with their experimental counterparts ($r(\ce{N-O}$)$\sim$\SI{1.514}{\angstrom} for NO~\cite{nist2005}; r(N-O)$\sim$\SI{1.193}{\angstrom}; $\theta$(O-N-O)$\sim$\SI{134.1}{\degree} for \ce{NO2}~\cite{herz66}).

\begin{figure}
\includegraphics[width=\linewidth, trim={0cm 0cm 0cm 1.3cm },clip]{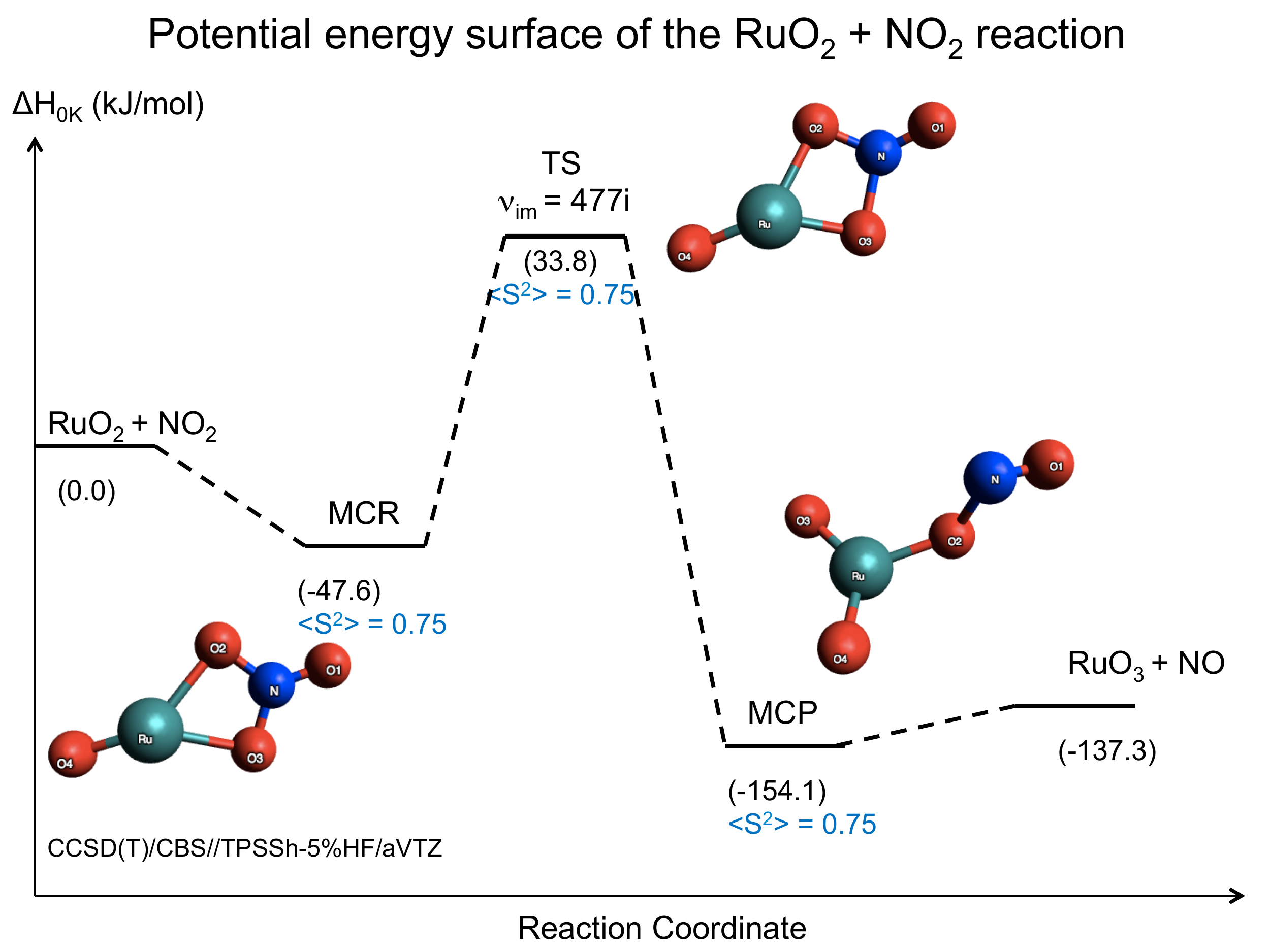}
\caption{\small Reaction coordinate at \SI{0}{\kelvin} calculated at the CCSD(T)/CBS//TPSSh-5\%HF/aVTZ level of theory for reaction \ce{RuO2 + NO2 -> RuO3 + NO} with schematic representations of the intermediate species involved.}
\label{fig:reacNo2r3}
\end{figure}

\begin{figure}
\includegraphics[width=\linewidth, trim={0cm 0cm 0cm 1.3cm },clip]{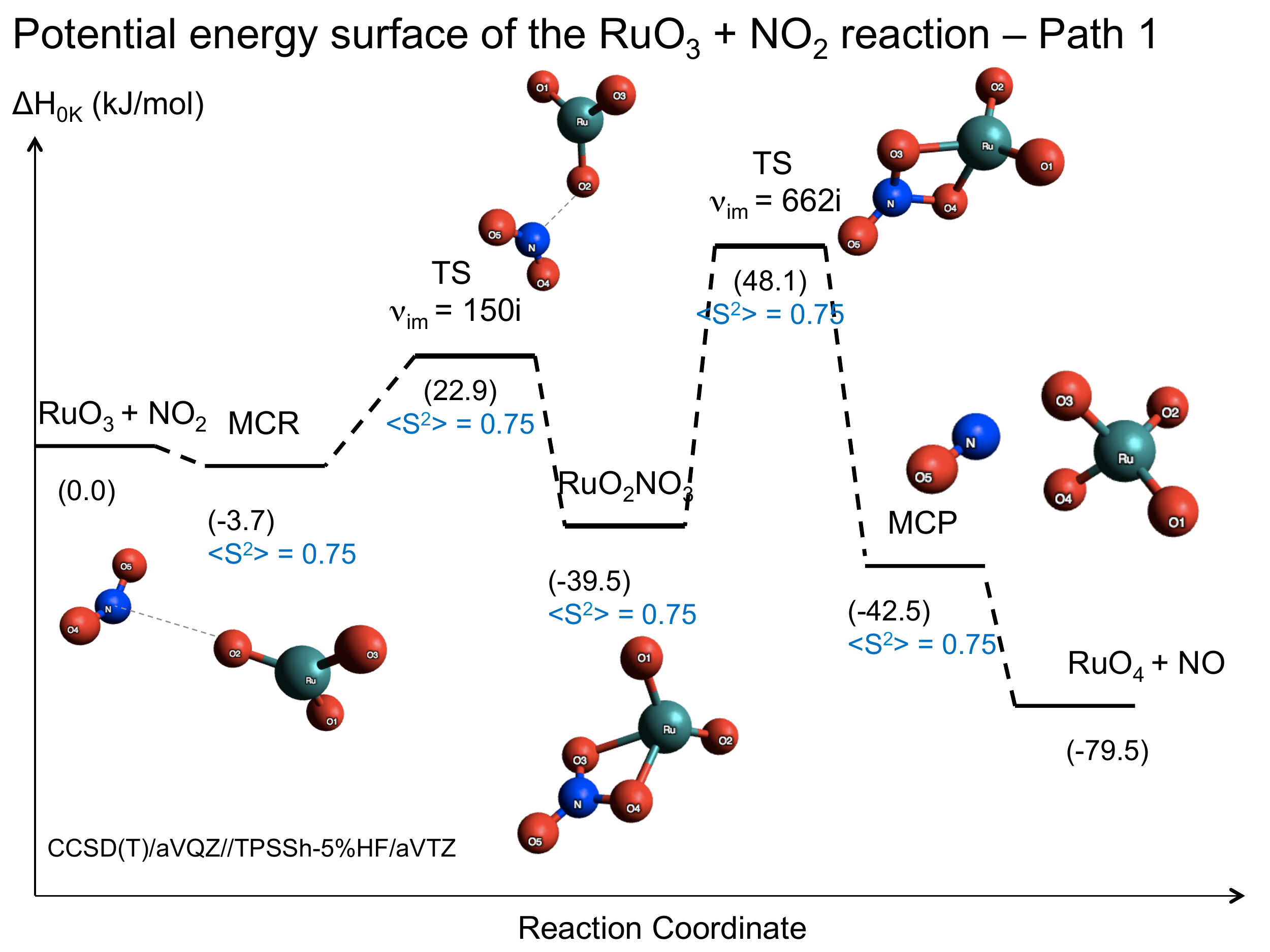}
\caption{\small Reaction coordinate at \SI{0}{\kelvin} calculated at the CCSD(T)/CBS//TPSSh-5\%HF/aVTZ level of theory for path~1 in reaction \ce{RuO3 + NO2 -> RuO4 + NO}, with schematic representations of the intermediate species involved.}
\label{fig:reacNo2r4-p1}
\end{figure}

\begin{figure}
\includegraphics[width=\linewidth, trim={0cm 0cm 0cm 1.3cm },clip]{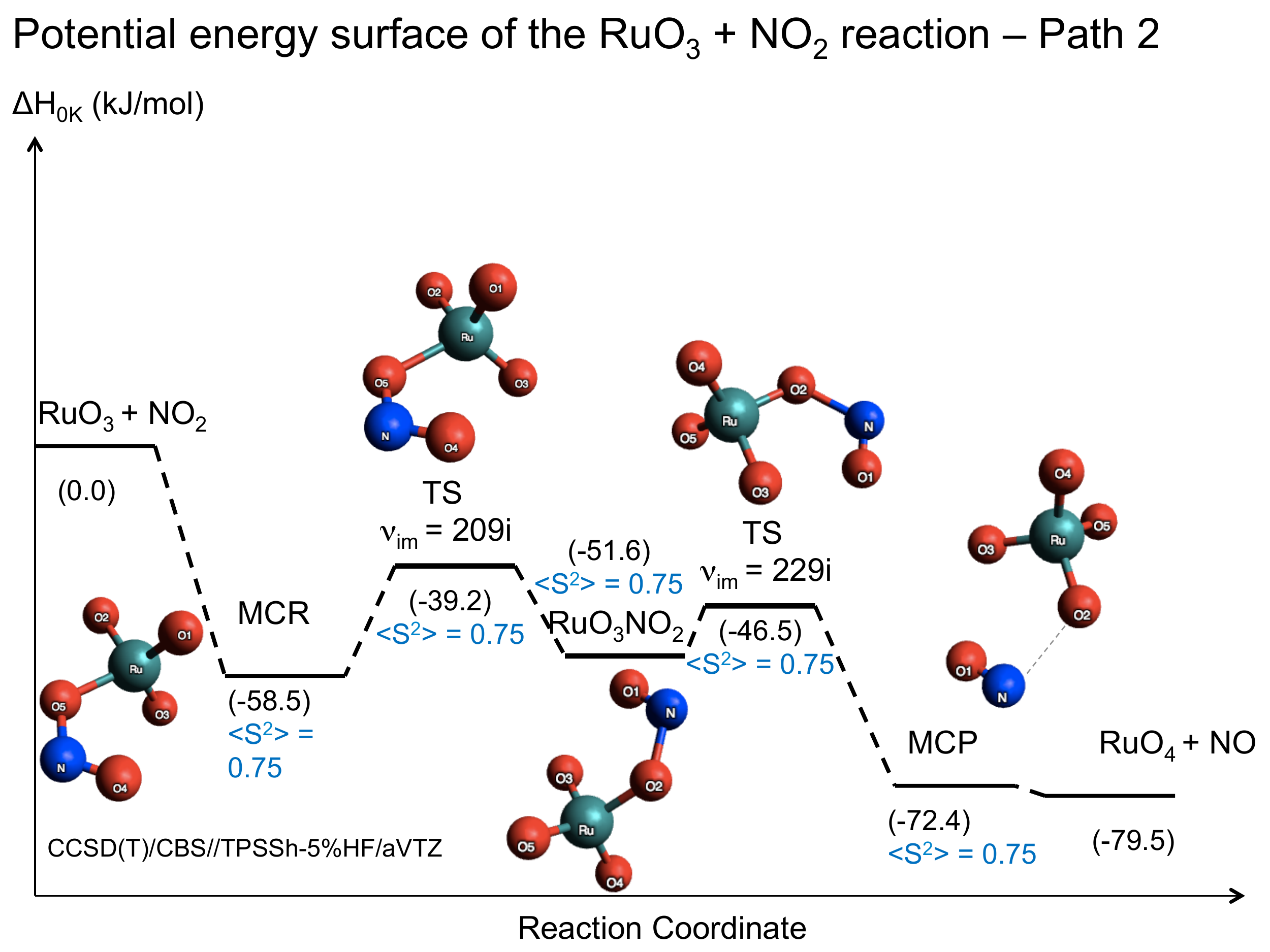}
\caption{\small Reaction coordinate at \SI{0}{\kelvin} calculated at the CCSD(T)/CBS//TPSSh-5\%HF/aVTZ level of theory for path~2 in reaction \ce{RuO3 + NO2 -> RuO4 + NO}, with schematic representations of the intermediate species involved.}
\label{fig:reacNo2r4-p2}
\end{figure}

The structures of intermediate species involved in the formation of~{\Rc} in reaction~\ref{eqru-noa} and relative enthalpies at \SI{0}{\kelvin} are shown in Fig.~\ref{fig:reacNo2r3}. For the formation of~{\Rd}, two reaction mechanisms, noted paths 1 and 2, were explored and are illustrated in Fig.~\ref{fig:reacNo2r4-p1} and Fig.~\ref{fig:reacNo2r4-p2}, respectively. Path~1 showcases nitrogen oxide forming a bond with one of ruthenium trioxide. Path~2 investigates Ru element forming a bond directly with the oxygen of nitrogen oxide. The corresponding geometrical parameters are reported in Tables~\ref{tab:runxo-react1a}, ~\ref{tab:runxo-react1b-path1}, and ~\ref{tab:runxo-react1b-path2}.

\begin{table}[t]
\centering
\caption{\small Structural parameters (bond lengths $r$ in~\si{\angstrom} ) imaginary vibrational frequency (\si{\per\cm}), and ZPE (\kjoule), for the transition state and molecular complexes calculated  at the TPSSh-5\%HF/aVTZ level of theory, involved in the reaction \ce{RuO2 + NO2  -> RuO3 +NO}}
\begin{tabular}{l*{3}{c}}
\toprule
Parameters &            {MCR(\ref{eqru-noa})}            &  {TS(\ref{eqru-noa})}         & {MCP(\ref{eqru-noa})} \\
\midrule
$r(\ce{N-O})$                     & 1.191 &     1.175      & 1.145  \\ 
$r(\ce{Ru-O_{NO_2}})$   & 1.316 &  \numrange{1.866}{1.988} &           \\
$r(\ce{Ru-O_{NO}})$        &           &                     & 1.693 \\
$r(\ce{Ru-O})$                   & 1.648          & 1.661  &  \numrange{1.673}{1.847}\\
$\theta(\ce{O-N-O})$     &  124.8            & 124.7 & 110.8 \\
$\theta(\ce{O-Ru-O_{NO_2}})$ & 93.0 & 72.9 & \\
$\theta(\ce{O-Ru-O_{NO}})$  &           &                     & 114.8\\
$\theta(\ce{O-Ru-O})$      &                       &   139.7  &  128.8\\
$\nu_{im}$                           &                   & {477\textbf{i}} & \\
ZPE &             46.8         & 39.8 &    38.2 \\
\bottomrule
\end{tabular}
\label{tab:runxo-react1a}
\end{table}
\begin{table*}
\centering
\caption{\small Structural parameters (bond lengths $r$ in~\si{\angstrom}), imaginary vibrational frequency (\si{\per\cm}), and ZPE (\kjoule), for the transition state and molecular complexes calculated at the TPSSh-5\%HF/aVTZ  level of theory, involved in path~1 for the reaction \ce{RuO3 + NO2  -> RuO4 +NO}}
\begin{tabular}{lccccc}
\toprule
Parameters &            {MCR(\ref{eqru-nob}-P1)}            &   {TS1(\ref{eqru-nob}-P1)}              & {\ce{RuO2NO3}}      & {TS2(\ref{eqru-nob}-P1)}  & {MCP(\ref{eqru-nob}-P1)} \\
\midrule
$r(\ce{N-O})$                      & \numrange{1.197}{1.200}  & \numrange{1.184}{1.188} &   1.184    &  1.183 & 1.125   \\
$r(\ce{Ru-O_{NO_2}})$      & 4.008           & 3.523   &         2.109    &  1.910 &             \\
$r(\ce{Ru-O})$                    & 1.687   & 1.686 -1.727 &        1.689    &       1.693    &    \numrange{1.694}{1.735} \\
$r(\ce{N-O_{RuO_3}})$  & 2.893   & 2.13  &        1.134    &  \numrange{1.402}{1.483} &                  \\
$r(\ce{N-O_{RuO_4}})$  &   &   &     &   &       2.198     \\
$\theta(\ce{O-N-O})$       & 134.7 & 137.6 &            124.2                      &   125.3 &                \\
$\theta(\ce{O-Ru-O_{NO_2}})$ & 104.0 & 103.8 &      62.3           &   70.9    &              \\
$\theta(\ce{O_{RuO_4}-N-O_{RuO_4}})$  &    &          &        &              &     71.7                   \\
$\theta(\ce{O-Ru-O})$       & \numrange{119.8}{120.3} & \numrange{117.9}{123.4}  &   \numrange{102.6}{124.8}   & \numrange{109.7}{120.9}   & \numrange{95.7}{115.7}   \\
$\nu_{im}$              &                      & 150\textbf{i} &           &   662\textbf{i} &\\
ZPE                     & 44.6      & 47.2  & 54.2 & 48.0  &  \\
\bottomrule 
\end{tabular}
\label{tab:runxo-react1b-path1}
\end{table*}
\begin{table*}
\centering
\caption{\small Structural parameters (bond lengths r are~\si{\angstrom}) imaginary vibrational frequency (\si{\per\cm}), and ZPE (\kjoule), for the transition states and molecular complexes calculated at the TPSSh-5\%HF/aVTZ  level of theory, involved in path~2 for the reaction \ce{RuO3 + NO2  -> RuO4 +NO}}
\begin{tabular}{lccccc}
\toprule
Parameters &    {MCR(\ref{eqru-nob}-P2)}     &  {TS1(\ref{eqru-nob}-P2)}      & {\ce{RuO3NO2}}    & {TS2(\ref{eqru-nob}-P2)} & {MCP(\ref{eqru-nob}-P2)}\\
\midrule
$r(\ce{N-O})$                      & 1.178-1.421   & 1.204-1.313 &  1.133    &  1.126  & 1.137\\
$r(\ce{Ru-O_{NO_2}})$   &  1.931          &   2.028       &  1.843      &   &             \\
$r(\ce{O_{RuO_4}-N_{NO}})$      &                      &    &     1.668  &  1.796 & 2.358   \\
$r(\ce{Ru-O})$                    & 1.688-1.725   & 1.685-1.723 & 1.692-1.733  & 1.693-1.728  & 1.696-1.711\\
$\theta(\ce{O-N-O})$       &114.8  &  114.5 &    115.0      &   &                \\
$\theta(\ce{O-Ru-O_{NO_2}})$ & 109.3  & 115.9 &    100.6            &      &              \\
$\theta(\ce{Ru-O-N_{NO}})$  &    &          &   123.9     &      119.7        &     114.9                     \\
$\theta(\ce{O-Ru-O})$       & 109.7-121.5  & 108.9-120.9 &  108.0-120.1    &  107.7-116.7  & 106.7-110.7    \\
$\nu_{im}$              &       & 209\textbf{i} &           &   229\textbf{i} &\\
ZPE                     &  47.1 &  47.5 & 47.4 & 45.6  & 46.2\\
\bottomrule
\end{tabular}
\label{tab:runxo-react1b-path2}
\end{table*}


The investigation of reaction~\ref{eqru-noa} pathway was conducted by approaching \ce{NO2} species to \ce{RuO2} oxide. The most stable potential energy led to an MCR(\ref{eqru-noa}) complex adopting a ring-like structure in which {\Rb} and \ce{NO2} interact through two symmetric covalent bonds (\SI{1.316}{\angstrom}) between $\ce{O_{RuO_2}-N_{NO_2}}$, and $\ce{Ru-O_{NO_2}}$, forming a terminal nitrate bidentate like structure. However, the \ce{N-O} bond distance, \SI{1.191}{\angstrom} remains similar to its counterparts in nitrogen dioxide (\SI{1.199}{\angstrom}). Such observation is consistent with our previous work~\cite{mira15} which underlined competitive effects between an increase of oxide bond lengths and increase of oxide charges (in this case on O atom) which kept the oxide bonds similar. In addition, the analysis of spin density of MCR(\ref{eqru-noa}) (Fig.~\ref{fig:spin_3_mcr} of the ESI) compound shows that the spin density of \ce{NO2} is transferred to Ru oxide, acknowledging a chemisorbed like structure, consistent within Lee~{\etal} work~\cite{dix20}, who observed chemisorbed-like structures for group IV \ce{(MO2)_n}, M=Ti, Zr, Hf dioxides nanoclusters from DFT and CCSD(T) calculations.
The bond angle $\theta$(O-N-O) decreases by ca. \SI{9}{\degree}. These results suggest that the {\Rb} electrons are the ones mostly involved in these two symmetric bonds in this complex; this is confirmed by the fact that the \ce{Ru-O} bond lengths are shorter, \SI{1.648}{\angstrom}, and the $\theta(\ce{O-Ru-O})$ bond angle is more acute, \SI{140.2}{\degree}, than in the {\Rb} molecule.

The MCP associated with the formation of {\Rc} presents a \ce{Ru-O} bond lengths lying between \SIlist{1.692;1.847}{\angstrom}, and a $\theta$(O-Ru-O) bond angle of \SI{128.8}{\degree}, slightly larger than those found in {\Rc} product. On the contrary, the \ce{N-O} bond length \SI{0.368}{\angstrom} shorter than the \ce{NO} equilibrium distance. These geometrical changes are induced by the highly covalent bond between $\ce{O_{RuO_3}}$ and N atoms with a length of \SI{1.693}{\angstrom}, that stabilizes {\Rc} and \ce{NO} within MCP species. The analysis of the spin density reflected a chemisorbed like structure (Fig.\ref{fig:spin_3_mcp} of the ESI), within a monodentate \ce{NO2} terminal, similar to the formed \ce{(MO2)NO2} chemisorbed species in Lee~{\etal}'s~work~\cite{dix20}.

The TS(\ref{eqru-noa}) connecting the MCR and MCP was found with an imaginary frequency of 477i~\si{\per\cm}, within a terminal nitrate bidentate like structure. The pro and post-reactants are found energetically more stable than the associated reactants and products, as shown in Fig.~\ref{fig:reacNo2r3}: the MCR complex lies $\sim$~\SI{48}{\kjoule} below the reactants, while the MCP one is $\sim$~\SI{17}{\kjoule} below the products.
This confirms that the identified TS(\ref{eqru-noa}) is the transient species leading to the formation of {\Rc} and NO species from the reaction between {\Rb} and \ce{NO2} compounds. However, the TS barrier is larger by about \SI{20}{\kjoule} than the one in reaction~\ref{eqru-n2a}, with \ce{N2O} species. This implies that this reaction will be energetically less favorable at \SI{0}{\kelvin}. This finding is comforted by $\ce{(N-O)_{N_2O}}$ and $\ce{(N-O)_{NO_2}}$ binding energies of \SIlist{-167.63;-306.3}{\kjoule}, respectively, obtained using {\dEr} of \ce{NO2= NO + O} and \ce{N2O = N2 + O}, respectively, computed from the NIST standard values~\cite{nist2005}.

For the formation of~{\Rd} in reaction~\ref{eqru-nob}, the two reactions paths explored both involve two-step mechanisms. In the reaction path~1, the connection of MCR(\ref{eqru-nob}-P1) to intermediate \ce{RuO2NO3} species by TS1(\ref{eqru-nob}-P1) corresponds to the formation of a ring-like structure between N, \ce{O_{\ce{NO2}}}, Ru, and \ce{O_{\Rc}} atoms. The molecular complex \ce{RuO2NO3}, which involves a nitrogen trioxide combined with a ruthenium dioxide, reveals geometric parameters for \ce{N-O} bond length (\SI{1.198}{\angstrom}) and O-N-O bond angle (\SI{132.2}{\degree}), smaller than their counterparts in \ce{NO3} (\SI{1.238}{\angstrom} and \SI{120}{\degree}, respectively \cite{herz66}), and closer to those featured in \ce{NO2}.
The analysis of the MCR(\ref{eqru-nob}-P1) spin density reveals a physisorbed-like structure (See Fig.\ref{fig:spin_4_p1_mcr} of the ESI), with the spin density localized on \ce{NO2}, unlike the intermediate species \ce{RuO2NO3} (See Fig~\ref{fig:spin_4_p1_interm} of the ESI) in which the \ce{NO2} transfers its spin density to the Ru-oxide and binds to it in a monodentate fashion. A similar result was found for adsorbed \ce{NO2} species onto \ce{MO3} trioxide clusters, with M=Cr and Mo~\cite{dix20}. Although the MCP(\ref{eqru-nob}-P1) portrayed a physisorbed like structure, within a terminal monodentate \ce{NO2} structure (Fig.\ref{fig:spin_4_p2_mcp}), and a spin density localized onto the Ru metal oxide. 

In reaction path~2, the TS1(\ref{eqru-nob}-P2) ensures the bonding between \ce{O_{\ce{NO2}}} and the Ru metallic center of~{\Rc}, to form the intermediate species \ce{RuO3NO2}. The latter complex exhibits Ru-O and N-O bond lengths slightly longer to those computed in {\Rc} and \ce{NO2}. In MCR(\ref{eqru-nob}-P2), the spin density (See Fig.~\ref{fig:spin_4_p2_mcr} of the ESI) is localized on the Ru-O unit, differing from the physisorbed molecular reactant species of path-1. This chemisorbed character could explain that this reactive molecular complex species lies \SI{58.5}{\kjoule} below the reactant fragments, as well as the TS1(\ref{eqru-nob}-P2) barrier, which is the highest barrier in reaction path~2, but above the MCR(\ref{eqru-nob}-P2) by \SI{19.3}{\kjoule}. This transition state leads to the formation of the intermediate chemisorbed species \ce{RuO3NO2} (Fig.\ref{fig:spin_4_p2_interm}), within a terminal monodentate \ce{NO2} structure, slightly higher in energy than the MCR(\ref{eqru-nob}-P2).

For the second step of the reaction mechanism, the breaking of the symmetric bonds between \ce{O_{\Rc}-N}, and \ce{Ru-O_{\ce{NO2}}}, the shrinkage of \ce{Ru-O} bonds, and finally the elongation of \ce{Ru-N} bond distance, is ensured by TS2(\ref{eqru-nob}-P2), to form the \ce{RuO4\bond{...}NO} product complex. This MCP species portrayed a physisorbed like shape, as the spin density is localized on \ce{NO2} (See Fig.~\ref{fig:spin_4_p2_mcp} of the ESI). In addition, the MCP(\ref{eqru-nob}-P2) species has $\theta$(O-Ru-O )($\sim$\SI{108.7}{\degree}) and $r(\ce{Ru-O})$($\sim$\SI{1.702}{\angstrom}) values close to those of the {\Rd} product. This MCP is geometrically different from the one identified for the path~1 (MCP(\ref{eqru-nob}-P1)), which had larger $\theta$(O-Ru-O)($\sim$\SI{105.7}{\degree}) and r(Ru-O)($\sim$\SI{1.714}{\angstrom}) values. These geometrical characteristics are corroborated by the computed relative energies at \SI{0}{\kelvin} of the MCPs, that lie \SIlist{37; 7.1}{\kjoule} above the products for MCP(\ref{eqru-nob}-P1) and MCP(\ref{eqru-nob}-P2), respectively.
It should be noticed that the MCP(\ref{eqru-nob}-P2) relative energy of -72.4~{\kjoule} is in the range of calculated adsorption enthalpies from Lee~{\etal}'s work~\cite{dix20} for group VI metal trioxide nanoclusters (from -206.8~{\kjoule} (\ce{(CriO3)NO2}, terminal bidentate nitrate like structure) to -70.3~{\kjoule}(\ce{(CriO3)NO2}, terminal bidentate \ce{NO2} like structure), comforting the appropriateness of proposed reaction pathway. 
In addition, the highest TS2(\ref{eqru-nob}-P1) barrier has a relative energy with respect to the reactants equal to \SI{48.1}{\kjoule}, twice as small as those determined in reaction (\ref{eqru-n2b}).

In conclusion, the reaction (\ref{eqru-nob}) path~2 is the most energetically favorable one to form~{\Rd}, in contrast to the thermodynamic calculations of the reaction Gibbs free energies~\cite{miradji-thesis}, which suggested that the reaction with nitrous oxide should be more spontaneous than the one with nitrogen oxide. The kinetic parameters of these reaction pathways are investigated in following section.

\subsection{Kinetic Parameters}

The calculations of the temperature dependence of the rate constants have been performed at the CCSD(T)/CBS//TPSSh-5\%HF/aVTZ  for the reactions with \ce{N2O} and \ce{NO2}, whose values are reported in Table~\ref{tab:are-tab}. For the formation of~{\Rd} by \ce{NO2} oxidation of {\Rc}, we considered that the formation of the molecular complex reactant will be the most limiting step, especially with increasing temperature. However, in our temperature range of interest it is likely that the MCR will be formed thus we approximate ~{\Rd} as a fast reaction.
The rate constants for \ce{N2O} and \ce{NO2} oxidation processes were fitted with the Arrhenius equation: 
\begin{align}
\label{eq:Arrhenius}
k(T) = B\times T^{n}\exp(-E_a/RT),
\end{align}
where $R$ is the gas constant and $T$ is the temperature. The Arrhenius parameters are the activation energy $E_a$, the pre-exponential factor, $B$, and unit less $n$. The Arrhenius parameters adjusted to eq.~\ref{eq:Arrhenius} are given in Table~\ref{tab:are-tab2}.
\begin{table*}
\caption{Rate constants in \si{\kunit}, calculated at the CCSD(T)/CBS//TPSSh-5\%HF/aVTZ level of theory.}
\scalebox{0.90}{
\begin{tabular}{l*8{S[table-format = 2.2e2]}}
\toprule	
Reactions & \multicolumn{8}{c}{Temperature (\si{\kelvin})}\\
\cmidrule(l){2-9} 
& {250} & {300} & {400} & {600} & {800} & {1000} & {1300} & {1500} \\
\midrule
\textbf{Formation of~{\Rc}}\\
\ce{RuO2 + N2O -> RuO3 + N2}   & 3.50e-16 & 1.08e-15 & 5.30e-15 & 3.75e-14 & 1.29e-13 & 3.12e-13 & 8.28e-13 &  1.37e-12 \\
\ce{RuO2 + NO2 -> RuO3 + NO} & 2.22e-22 & 3.53e-21 & 6.20e-20 & 6.79e-18 & 6.40e-17 & 2.88e-16 & 1.38e-15 & 3.00e-15 \\
\hline
\\
\textbf{Formation of~{\Rd}} \\
\ce{RuO3 + N2O -> RuO4 + N2} &1.07e-37 & 8.86e-34 & 8.16e-29 & 1.05e-23 & 4.77e-21 & 2.16e-19 & 8.48e-18  & 4.65e-17 \\
\bottomrule
\end{tabular}%
}
\label{tab:are-tab}
\end{table*}
The computed rate constants related to the formation of {\Rc} through the oxidation of~{\Rb} by \ce{N2O}, range from \num{1e-16} at \SI{250}{\kelvin} to  \SI{1e-12}{\kunit} at \SI{1500}{\kelvin}. 
The oxidation of~{\Rb} by  \ce{NO2} shows a similar temperature behavior, with values varying from \num{1e-22} to  \SI{1e-15}{\kunit}. These two sets of results indicated that the mechanism involving the nitrogen dioxide is slightly slower than the one with the nitrous oxide, confirming the reactions coordinates curves depicted in Fig.~\ref{fig:reacNo2r3}, reflecting the fact that the TS barrier is larger by $\sim$\SI{20}{\kjoule} in reaction~\ref{eqru-noa} than in reaction~\ref{eqru-n2a}.

\begin{table*}
\caption{Arrhenius parameters calculated over the temperature range \SIrange{250}{2500}{\kelvin} from energy profiles calculated at the CCSD(T)/CBS//TPSSh-5\%HF/aVTZ level of theory.}
\centering
\begin{tabular}{l*{1}{S[table-format = 2.2e1]}*{2}{S}*{1}{S[table-format = 2.2e1]}}
\toprule
& {B}& {$n$}  & {$E_a$}  & {$k(\SI{298}{\kelvin})$} \\
& {(\si{\kunit})} & & {(\si\kjoule)} & {(\si{\kunit})}\\
\midrule
\multicolumn{5}{l}{Formation of {\Rc}}\\
\ce{RuO2 + N2O -> RuO3 + N2} & 2.21e-21  & 2.85 & 8.0  & 9.79e-16\\
\ce{RuO2 + NO2 -> RuO3 + NO} &4.01e-24 &3.09 & 27.1 & 3.22e-21\\
\midrule
\multicolumn{5}{l}{Formation of {\Rd}}\\
\ce{RuO3 + N2O -> RuO4 + N2}         & 5.23e-22   & 2.72   & 106.4 & 6.14e-34\\
\bottomrule
\end{tabular}
\label{tab:are-tab2}
\end{table*}

The reaction process involving the oxidation of~{\Rc} by \ce{N2O} (reaction~\ref{eqru-n2b}) appears quite slow at our temperature scale, with rate constant values varying from \num{1e-37} at \SI{250}{\kelvin} to \SI{1e-17}{\kunit} at \SI{1500}{\kelvin}. These values are consistent with the TS barrier displayed in Fig.~\ref{fig:reacN2or4}, and emphasizing the fact that the 
reaction~\ref{eqru-nob} involving nitrous oxide, path 2, is the more likely pathway to form \Rd, the kinetically limiting step might only reside from the MCR complex to overcome the TS1 barrier, as the MCR complex lies energetically below the reactants at 0K. 

\subsection{Theoretical Results Compared to Experimental Tests}

We can now discuss our theoretical results in the light of experimental data. In the framework of the VTT program~\cite{vtt14,kajan2016}, previously described in the introduction, the ruthenium transport under humid atmospheres, and humid atmospheres with air radiolysis products precursors, with different temperature gradients was studied. Table~\ref{tab:kajan-tab} summarizes the results reported in Kajan's thesis~\cite{kajan2016}.

\begin{table*}
\caption{Measurement  of transported Ru fraction in function of carrier gas and temperature in VTT tests~\cite{kajan2016}. The model primary circuit is made of either stainless steel tube or alumina tube samples. The temperature gradients range from \SIlist{1300;1500;1700}{\kelvin} down to ca. \SI{300}{\kelvin}}
\begin{tabular}{l*{3}{S}}
\toprule
Atmosphere ($T$) & {Ru released rate} &  {Transported \ce{RuO2_{(s)}}} & {Transported \ce{RuO4_{(g)}}}\\
&  {(\si{\mg\per\minute})} & {(\% Ru released)}&   {(\% Ru released)} \\
\midrule
Humid Air (\SI{1300}{\kelvin}) & 0.3 \pm 0.0  & 9.1 \pm 0.5  & 0.0 \pm0.0 \\
Humid Air (\SI{1500}{\kelvin}) & 3.2 \pm 0.2  & 12.8 \pm 0.6 & 0.0 \pm0.0 \\
Humid Air (\SI{1700}{\kelvin}) & 20.3 \pm 1.0  & 14.3 \pm 0.7 & 0.0 \pm0.0 \\
 &  &    & \\
Humid Air +  \ce{NO2} (\SI{1300}{\kelvin}) & 0.3 \pm 0.0    & 0.0 \pm 0.0 &  13.9 \pm0.7  \\
Humid Air +  \ce{NO2}  (\SI{1500}{\kelvin}) & 3.2 \pm 0.2   & 4.0 \pm 0.2  &  9.9 \pm0.5       \\
Humid Air +  \ce{NO2}  (\SI{1700}{\kelvin}) & 20.3 \pm 1.0 & 20.2 \pm 1.0   & 0.0 \pm0.0 \\
 &  &   & \\
Humid Air +  \ce{N2O} (\SI{1300}{\kelvin}) & 0.3 \pm 0.0  & 6.0 \pm 0.3 & 0.1 \pm0.0 \\
Humid Air +  \ce{N2O}  (\SI{1500}{\kelvin}) & 3.2 \pm 0.3   & 25.4 \pm 1.7 &  0.1 \pm0.0 \\ 
Humid Air +  \ce{N2O}  (\SI{1700}{\kelvin}) & 20.3 \pm 1.0  & 15.5 \pm 0.8\footnotemark[1] & 0.0 \pm0.0\footnotemark[1] \\
\bottomrule
\end{tabular}\\
\footnotemark[1]{measured for $T$ = \SI{1570}{\kelvin}}
\label{tab:kajan-tab}
\end{table*}

With an atmosphere containing 50 ppmV of \ce{NO2}, the transport of ruthenium tetroxide increased by 92\% at \SI{1300}{\kelvin} and 42\% at \SI{1500}{\kelvin}, by comparison to humid air atmospheres. The increase of {\Rd} fractions was attributed to the reaction between \ce{NO2} and {\Rc}, as expressed: 
% in equation~\ref{eq:tono}: 
\begin{equation}
\ce{RuO3 + NO2 -> RuO4 + NO}.
\label{eq:tono}
\end{equation}
The equilibrium constants K$_{eq}$ calculated by Kajan~{\etal}~\cite{kajan2016} were derived using HSC 5.11 chemistry software~\cite{hsc2002}, equal to  \numlist{28.55; 16.85; 11.3} at \SIlist{1300; 1500; 1700}{\kelvin}, respectively.

Our equilibrium constants obtained from the kinetic reactions rate constants calculations  appeared in good agreement with the ones derived from Kajan work's, though they come out slightly lower, \numlist{23.64; 9.37; 4.66} at \SIlist{1300; 1500; 1700}{\kelvin}, respectively. These differences could be explained by the small deviations at higher temperatures of our derived thermodynamic properties for Ru oxides, as discussed in our previous work~\cite{mira16}.
The decreasing amount of detected gaseous fraction of Ru as temperature increases can be attributed to the decomposition of \ce{NO2} with temperature~\cite{pola56,Huffman59}. This is fully supported by the free-energy calculations obtained for the formation of {\Rd} through nitrogen oxidation that shows higher values when temperature increases. 
At this stage, it is not clear how the decrease of aerosol formation is linked to the reaction of nitrogen oxide to form {\Rd}. Such aspects are investigated through the study of the nucleation process of Ru dimer in a parallel work~\cite{artiRudim2021}.

With an atmosphere containing 50 ppmV of \ce{N2O}, the experimental tests showcased a similar production of~{\Rd} gaseous fraction in comparison to pure humid air atmosphere, along with a slight increase as the temperatures rise.
These observations are also consistent with our quantum chemical data, as the reaction barrier in the oxidation process of {\Rc} by \ce{N2O} to form~{\Rd} is large. 

Altogether, this discussion leads us to conclude that the calculated kinetic rates for the formation of~{\Rd} in air radiolysis products atmospheres are fully consistent with experimental observations, even if some other phenomena can play a role like surface interactions.

\section{Conclusions}

The mechanisms and kinetics of the chemical reactions leading to the formation of~{\Rc} and~{\Rd} gaseous species under severe accident (SA) conditions of a nuclear power plant (NPP) were elucidated by state-of-the art quantum chemical approaches. An in-depth investigation of the reaction pathways involving two air radiolysis products \ce{N2O} and \ce{NO2} to form~{\Rd} and~{\Rc}, following experimental observations, was conducted. The coupled-cluster theory was then employed to compute the potential energies. 
The transition states obtained for the formation of~{\Rd} from \ce{NO2} oxidation appeared energetically lower than the reactants.

The derivation of the related kinetic rates to form~{\Rd} through the nitroxide species revealed that nitrogen oxide process is a faster mechanism than the one involving nitrous oxide, thus contrasting the thermodynamic predictions.
These results are consistent with the experimental observations and measurements acquired by the VTT program, which concluded to an increase of transported Ru gaseous fraction in humid atmospheres with air radiolysis precursors.

%Further investigations of the impact of the computed kinetics to model experimental measurements will be conducted in the near future to enhance the simulation tools to estimate Ru source term in nuclear severe accidents conditions.

%\begin{acknowledgements}
\section*{Acknowledgements}
The computer time for part of the theoretical calculations was kindly provided by the Centre R\'egional 
Informatique et d'Applications Num\'eriques de Normandie (CRIANN) of the University of Lille, and the HPC resources from GENCI-cines (grant 2015-project number~x2015086731). This work has been supported by  the French government through the Program "Investissement d'avenir" (LABEX CaPPA / ANR-11-LABX-0005-01 and I-SITE ULNE / ANR-16-IDEX-0004 ULNE), as well as by the Ministry of Higher Education and Research, Hauts de France council and European Regional Development Fund (ERDF) through the Contrat de Projets État-Région (CPER CLIMIBIO). The authors declare no competing financial interest.
%\end{acknowledgements}

\section*{Author contributions}
Dr. Faoulat Miradji: conceptualization, methodology, quantum chemical calculations, kinetics calculations, data analysis, data validation, writing (original draft), writing (review \& editing), visualization. Dr. Sidi M. O. Souvi: conceptualization, methodology, data analysis, data validation, writing (review \& editing). Dr. Laurent Cantrel: conceptualization, methodology, data analysis, data validation writing (review \& editing). Dr. Florent Louis: conceptualization, methodology, data analysis, data validation, writing (review \& editing). Dr. Val\'erie Vallet: conceptualization, methodology, quantum chemical calculations, data analysis, data validation, writing (original draft), writing (review \& editing).


\bibliography{BIB-tot}

\clearpage



\section*{Supplementary material}

# Supporting Information:

# Reactivity of Ru oxides with air radiolysis products investigated by theoretical calculations

Faoulat Miradji,[*,†,‡,¶,§] Sidi M. O. Souvi,[†,¶] Laurent Cantrel,[†,¶] Florent Louis,[‡,¶]

and Valérie Vallet[§]

[†]*Institut de Radioprotection et de Sûreté Nucléaire (IRSN), PSN-RES, Cadarache, St Paul Lez Durance, 13115, France*

[‡]*Univ. Lille, CNRS, UMR 8522 - PC2A - Physicochimie des Processus de Combustion et de l'Atmosphère, F-59000 Lille, France*

[¶]*Laboratoire de Recherche Commun IRSN-CNRS-Lille1 Cinétique Chimique, Combustion, Réactivité (C³R), Cadarache, Saint Paul Lez Durance,13115, France*

[§]*Univ. Lille, CNRS, UMR 8523 - PhLAM - Physique des Lasers Atomes et Molécules, F-59000 Lille, France*

E-mail: fmiradji@protonmail.com

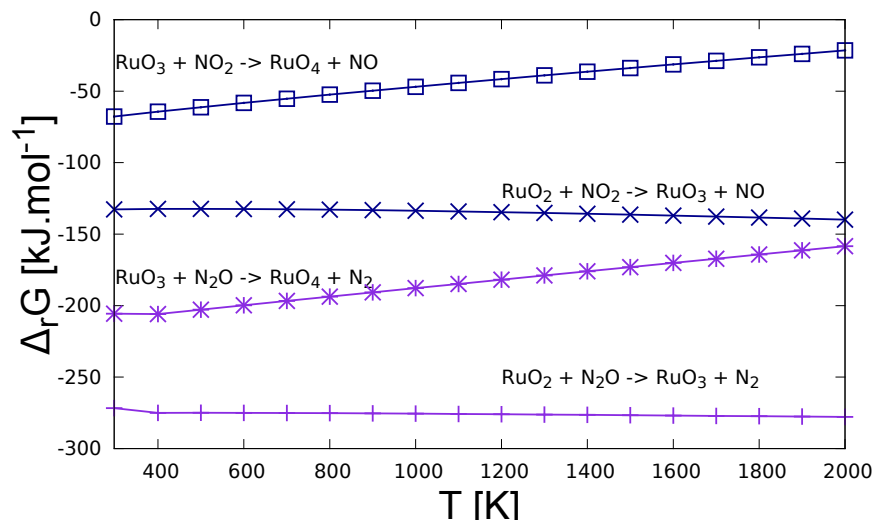

Figure S1: Gibbs energies of reaction curves of Ru oxides reactivity within $N_2O$ and $NO_2$ species

Table S1: Cartesian coordinates, rotational constant, zero point energy corrections (ZPE), point group, symmetry number and frequencies ($\omega$) of reactants and products.

| Properties | RuO$_2$ ($^1$A) | | | RuO$_3$ ($^1$A) | | | RuO$_4$ ($^1$A) | | |
|---|---|---|---|---|---|---|---|---|---|
| ZPE (kJ/mol) | 12.73 | | | 21.28 | | | 33.46 | | |
| Rotational constants (GHZ) | X | Y | Z | X | Y | Z | X | Y | Z |
|  | 108.0097558 | 5.9832979 | 5.6692450 | 7.4072384 | 7.4072384 | 3.7036192 | 4.1428122 | 4.1428122 | 4.1428122 |
| Full Point Group | C$_2$V | | | D$_3$H | | | T$_d$ | | |
| Symmetry number | 4 | | | 12 | | | 24 | | |
| COORDINATES | X | Y | Z | X | Y | Z | X | Y | Z |
| O |  |  |  | 0.000000 | 0.000000 | 0.000000 | 0.976394 | 0.976394 | 0.976394 |
| Ru | 0.000000 | 0.000000 | 0.116903 | 0.000000 | 1.686335 | 0.000000 | 0.000000 | 0.000000 | 0.000000 |
| O | 0.000000 | 1.624921 | -0.321483 | 1.460409 | -0.843167 | 0.000000 | -0.976394 | -0.976394 | 0.976394 |
| O | 0.000000 | -1.624921 | -0.321483 | -1.460409 | -0.843167 | 0.000000 | -0.976394 | 0.976394 | -0.976394 |
| O |  |  |  |  |  |  | 0.976394 | -0.976394 | -0.976394 |
| $\omega_e$, cm$^{-1}$ | 191.6001 | 962.6781 | 972.3287 | 66.2133 | 299.4090 | 299.4108 | 329.4897 | 329.4897 | 352.8578 |
|  |  |  |  | 962.6525 | 962.6542 | 964.6298 | 352.8578 | 352.8578 | 952.782 |
|  |  |  |  |  |  |  | 974.7355 | 974.7355 | 974.7355 |

| Properties | NO$_2$ ($^2$A$_1$) | | | NO ($^2\Sigma$) | | | N$_2$O ($^1\Sigma$) | | | N$_2$ ($^1\Sigma_g$) | | |
|---|---|---|---|---|---|---|---|---|---|---|---|---|
| ZPE (kJ/mol) | 22.45 | | | 11.47 | | | 28.77 | | | 14.31 | | |
| Rotational constants (GHZ) | X | Y | Z | X | Y | Z | X | Y | Z | X | Y | Z |
|  | 237.9974209 | 12.9477411 | 12.2796908 | 0.0000000 | 50.9464437 | 50.9464437 | 0.0000000 | 12.6296066 | 12.6296066 | 0.0000000 | 59.9838011 | 59.9838011 |
| Full Point Group | C2V | | | C1 | | | C*$_v$ | | | D*$_h$ | | |
| Symmetry number | 4 | | | 1 | | | 4 | | | 8 | | |
| COORDINATES | X | Y | Z | X | Y | Z | X | Y | Z | X | Y | Z |
| N |  |  |  |  |  |  | 0.000000 | 0.000000 | -0.071636 | 0.000000 | 0.000000 | 0.548485 |
| O | 0.000000 | 1.104602 | -0.142109 |  |  |  |  |  |  |  |  |  |
| N | 0.000000 | -1.104602 | -0.142109 | 0.000000 | 0.000000 | -0.614743 | 0.000000 | 0.000000 | -1.199636 | 0.000000 | 0.000000 | -0.548485 |
| O | 0.000000 | 0.000000 | 0.324820 | 0.000000 | 0.000000 | 0.537900 | 0.000000 | 0.000000 | 1.112364 |  |  |  |
| $\omega_e$, cm$^{-1}$ | 750.8319 | 1350.6832 | 1652.2626 | 1917.78 | | | 602.8360 | 602.8360 | 1308.3437 | 2392.94 | | |
|  |  |  |  |  |  |  | 2296.36 | | | | | |

Table S2: Cartesian coordinates, rotational constant, zero point energy corrections (ZPE), point group, symmetry number and frequencies ($\omega$) of intermediates complexes in the reaction (1) $RuO_2 + N_2O \longrightarrow RuO_3 + N_2$

| Properties | MCR(1) | | | TS (1) | | | MCP(1) | | |
|---|---|---|---|---|---|---|---|---|---|
| ZPE (kJ/mol) | 43.94 | | | 37.65 | | | 36.18 | | |
| Rotational constants (GHZ) | X | Y | Z | X | Y | Z | X | Y | Z |
| | 5.1297070 | 1.5161829 | 1.2821402 | 5.9809808 | 1.3966668 | 1.1766445 | 4.5828832 | 0.8077300 | 0.7755076 |
| Full Point Group | C1 | | | C1 | | | C1 | | |
| Symmetry number | 1 | | | 1 | | | 1 | | |
| COORDINATES | X | Y | Z | X | Y | Z | X | Y | Z |
| O | 1.176175 | 1.577971 | -0.208059 | 1.281787 | 1.542156 | -0.171247 | 0.266574 | 1.447628 | 0.712655 |
| Ru | 0.647602 | -0.000047 | 0.007961 | 0.653985 | -0.000002 | 0.018027 | 0.821082 | 0.000086 | 0.052381 |
| O | 1.175012 | -1.578527 | -0.207531 | 1.281754 | -1.542162 | -0.171336 | 1.910688 | 0.023576 | -1.232669 |
| O | -1.288350 | 0.000726 | 0.872462 | -1.184156 | -0.000002 | 0.618028 | 0.293349 | -1.470941 | 0.681997 |
| N | -2.185310 | 0.000247 | 0.040244 | -2.281084 | 0.000019 | -0.170186 | -3.448605 | -0.00334 | -0.328855 |
| N | -3.100000 | -0.000145 | -0.612423 | -3.406118 | 0.000000 | -0.257920 | -4.536038 | 0.002499 | -0.185519 |
| $\omega_e$, cm$^{-1}$ | 46.4335 | 60.8045 | 135.9801 | i590.0716 | 53.1052 | 105.0999 | 9.7277 | 11.3050 | 14.6701 |
| | 198.0938 | 237.3583 | 278.0440 | 180.1901 | 245.1330 | 275.4357 | 27.1644 | 31.7603 | 74.1334 |
| | 518.1889 | 552.8043 | 955.645 | 374.5826 | 420.0309 | 656.8689 | 299.9844 | 300.3471 | 962.0911 |
| | 986.0903 | 1127.8936 | 2248.4029 | 963.0379 | 986.6513 | 2034.3637 | 962.1891 | 964.0729 | 2392.3267 |

Table S3: Cartesian coordinates, rotational constant, zero point energy corrections (ZPE), point group, symmetry number and frequencies ($\omega$) of intermediates complexes in the reaction (2) $RuO_3 + N_2O \longrightarrow RuO_4 + N_2$

| Properties | MCR (2) | | | TS (2) | | | MCP (2) | | |
|---|---|---|---|---|---|---|---|---|---|
| ZPE (kJ/mol) | 50.73 | | | 49.96 | | | 48.62 | | |
| Rotational constants (GHZ) | X | Y | Z | X | Y | Z | X | Y | Z |
| | 4.6606935 | 0.5272251 | 0.5124401 | 4.5094315 | 1.2883910 | 1.1932432 | 4.1462380 | 0.8012075 | 0.8012060 |
| Full Point Group | C1 | | | C1 | | | C1 | | |
| Symmetry number | 1 | | | 1 | | | 1 | | |
| COORDINATES | X | Y | Z | X | Y | Z | X | Y | Z |
| O | -2.077751 | -1.447406 | 0.081334 | -1.730201 | 1.191582 | -0.502419 | -2.399824 | 0.006821 | -0.001810 |
| Ru | -1.222062 | -0.000131 | -0.018906 | -0.631037 | -0.002725 | -0.035775 | -0.710126 | 0.000055 | -0.000024 |
| O | 0.454026 | -0.022183 | -0.199432 | -1.237449 | -1.498666 | 0.456301 | -0.143324 | 1.525341 | -0.454124 |
| O | -2.041450 | 1.469189 | 0.061947 | 1.231654 | -0.522109 | -0.863505 | -0.153057 | -0.372177 | 1.550811 |
| N | 3.786327 | -0.000212 | -0.013338 | 2.174776 | -0.057830 | -0.165012 | 3.311796 | -0.002784 | 0.001148 |
| N | 3.944403 | -0.010339 | -1.131918 | 3.217841 | 0.215200 | 0.141581 | 4.408568 | 0.002171 | -0.000901 |
| O | 3.622126 | 0.010353 | 1.162235 | 0.488162 | 0.706482 | 1.126890 | -0.153419 | -1.159753 | -1.094963 |
| $\omega_e$, cm$^{-1}$ | 5.1034 | 9.5575 | 12.2012 | i572.7340 | 73.7574 | 135.1986 | 10.1784 | 14.4004 | 29.4317 |
| | 25.8735 | 43.9646 | 82.1165 | 172.2392 | 222.7866 | 269.9064 | 45.0546 | 45.4828 | 330.6914 |
| | 299.3357 | 300.4401 | 602.4362 | 293.8226 | 358.2030 | 482.2226 | 330.8559 | 353.4842 | 354.8703 |
| | 03.4794 | 961.1501 | 963.0181 | 573.3981 | 810.8662 | 885.4292 | 354.9362 | 950.0959 | 971.5989 |
| | 964.9427 | 1308.7591 | 2298.3460 | 938.0239 | 942.6752 | 2194.1663 | 972.3110 | 972.3514 | 2393.2102 |

Table S4: Cartesian coordinates, rotational constant, zero point energy corrections (ZPE), point group, symmetry number and frequencies ($\omega$) of intermediates complexes in the reaction (3) $RuO_2 + NO_2 \longrightarrow RuO_3 + NO$

| Properties | MCR (3) | | | TS (3) | | | MCP (3) | | |
|---|---|---|---|---|---|---|---|---|---|
| ZPE (kJ/mol) | 46.82 | | | 39.78 | | | 38.21 | | |
| Rotational constants (GHZ) | X | Y | Z | X | Y | Z | X | Y | Z |
| | 12.1247530 | 1.4050487 | 1.2867721 | 10.5644333 | 1.5096537 | 1.3601817 | 5.9117491 | 1.3365820 | 1.1500726 |
| Full Point Group | C1 | | | C1 | | | | | |
| Symmetry number | 1 | | | 1 | | | | | |
| COORDINATES | X | Y | Z | X | Y | Z | X | Y | Z |
| O | 2.935606 | -0.005089 | 0.012196 | 2.857601 | -0.016436 | -0.205770 | 3.287999 | 0.000071 | 0.018507 |
| N | 1.747156 | -0.001991 | 0.093568 | 1.782548 | -0.049026 | 0.266331 | 2.254462 | -0.000081 | 0.511212 |
| O | 0.992632 | -1.080686 | 0.097267 | 0.924918 | -1.101091 | 0.169973 | 1.115294 | -0.000136 | -0.577604 |
| Ru | -0.751688 | 0.003184 | -0.123102 | -0.688071 | 0.024759 | -0.123198 | -0.681606 | 0.000008 | -0.080058 |
| O | 0.997828 | 1.080912 | 0.103308 | 0.734797 | 1.183769 | 0.214846 | -1.313479 | 1.514834 | 0.276031 |
| O | -2.320544 | -0.010906 | 0.382418 | -2.292655 | -0.159520 | 0.265502 | -1.313632 | -1.514741 | 0.276075 |
| $\omega_e$, cm$^{-1}$ | 103.2030 | 122.8006 | 171.3540 | *i477.1376* | 107.9314 | 163.6946 | 50.4747 | 104.4357 | 107.0845 |
| | 341.4383 | 367.0201 | 638.2001 | 209.5668 | 303.5702 | 429.5794 | 224.6151 | 225.7755 | 270.4422 |
| | 716.0601 | 750.6165 | 922.9649 | 551.3521 | 668.4614 | 755.6998 | 343.0228 | 518.8752 | 791.6028 |
| | 982.2243 | 1088.3080 | 1622.6906 | 858.1584 | 964.7844 | 1637.5644 | 955.928 | 977.4133 | 1818.5923 |

Table S5: Cartesian coordinates, rotational constant, zero point energy corrections (ZPE), point group, symmetry number and frequencies ($\omega$) of intermediates complexes in the reaction (4) $RuO_3 + NO_2 \longrightarrow RuO_4 + NO$, reaction path 1

| Properties | MCR (4-P1) | | | TS1 (4-P1) | | | MCP1 (4-P1) | | |
|---|---|---|---|---|---|---|---|---|---|
| ZPE (kJ/mol) | 44.60 | | | 47.24 | | | 53.15 | | |
| Rotational constants (GHZ) | X | Y | Z | X | Y | Z | X | Y | Z |
| | 4.6244958 | 0.4907772 | 0.4763704 | 4.1110252 | 0.8753627 | 0.8486907 | 4.1933030 | 1.1058445 | 1.0370194 |
| Full Point Group | C1 | | | C1 | | | C1 | | |
| Symmetry number | 1 | | | 1 | | | 1 | | |
| COORDINATES | X | Y | Z | X | Y | Z | X | Y | Z |
| O | 2.227767 | 1.402889 | -0.015836 | 1.590395 | 1.484945 | 0.362296 | 1.397496 | 1.536232 | 0.292272 |
| Ru | 1.293919 | -0.000626 | -0.000652 | 0.944430 | 0.000005 | -0.107691 | 0.862346 | -0.000740 | -0.120983 |
| O | -0.390159 | 0.106043 | -0.007680 | -0.582562 | -0.00002 | -0.91454 | -0.883986 | -0.003585 | -0.836214 |
| O | 2.043702 | -1.510287 | 0.022653 | 1.590469 | -1.484920 | 0.362244 | 1.395960 | -1.534744 | 0.305205 |
| N | -3.762405 | 0.320465 | -0.006566 | -2.450455 | -0.000026 | 0.125042 | -2.048148 | 0.001061 | 0.105676 |
| O | -3.85663 | -0.165176 | -1.099835 | -3.364372 | 0.000043 | -0.628906 | -3.098695 | -0.005262 | -0.46315 |
| O | -3.849133 | -0.110434 | 1.110028 | -2.284144 | -0.000054 | 1.301795 | -1.761548 | 0.010502 | 1.274826 |
| $\omega_e$, cm$^{-1}$ | 7.4485 | 8.2264 | 9.7245 | *i150.9221* | 32.6407 | 57.5459 | 67.1536 | 84.1349 | 86.9353 |
| | 25.0685 | 37.8068 | 53.0252 | 62.8187 | 135.4425 | 231.5574 | 137.1682 | 244.0293 | 271.9447 |
| | 81.0845 | 298.0720 | 299.4307 | 261.9617 | 293.2727 | 322.2995 | 382.5503 | 468.3788 | 673.2463 |
| | 749.7321 | 962.3188 | 964.4556 | 717.9817 | 783.0545 | 955.1125 | 734.5048 | 797.5590 | 969.0069 |
| | 965.5071 | 1346.5539 | 1648.1523 | 961.6475 | 1309.7889 | 1772.9950 | 974.7986 | 1284.6279 | 1709.4752 |

Table S6: Cartesian coordinates, rotational constant, zero point energy corrections (ZPE), point group, symmetry number and frequencies ($\omega$) of intermediates complexes in the reaction (4) $RuO_3 + NO_2 \longrightarrow RuO_4 + NO$, reaction path 1

| Properties | $RuO_2NO_3$ (4-P1) | | | TS2 (4-P1) | | | MCP (4-P1) | | |
|---|---|---|---|---|---|---|---|---|---|
| ZPE (kJ/mol) | 54.16 | | | 48.00 | | | 48.32 | | |
| Rotational constants (GHZ) | X | Y | Z | X | Y | Z | X | Y | Z |
| | 4.7439577 | 1.3016302 | 1.1812363 | 4.5756841 | 1.3969871 | 1.2845363 | 4.0632015 | 1.2885707 | 1.2304325 |
| Full Point Group | C1 | | | C1 | | | C1 | | |
| Symmetry number | 1 | | | 1 | | | 1 | | |
| COORDINATES | X | Y | Z | X | Y | Z | X | Y | Z |
| O | -1.020173 | 1.622270 | -0.339646 | 1.314760 | 1.563326 | 0.046565 | 1.264987 | 1.570622 | -0.000742 |
| Ru | -0.716661 | -0.016481 | -0.012721 | 0.631369 | 0.015837 | 0.034445 | 0.588006 | 0.016582 | 0.000022 |
| O | -1.850655 | -1.197442 | 0.317747 | 1.604388 | -1.351952 | -0.187505 | 1.692354 | -1.268250 | -0.004887 |
| O | 1.144989 | 0.420638 | 1.005679 | -0.910537 | 0.000112 | -1.087939 | -0.567995 | -0.154289 | -1.283269 |
| N | 1.836528 | -0.086010 | 0.039374 | -1.838889 | -0.21623 | -0.05441 | -2.308117 | -0.498049 | 0.001044 |
| O | 1.038430 | -0.546600 | -0.928785 | -0.923399 | -0.302036 | 1.109564 | -0.559784 | -0.156771 | 1.289184 |
| O | 3.022080 | -0.132962 | -0.019484 | -2.948712 | 0.192645 | -0.022523 | -3.043994 | 0.353277 | -0.001320 |
| | 33.0293 | 118.7055 | 156.040 | i662.3057 | 92.3427 | 175.3793 | 70.2301 | 100.8289 | 109.6790 |
| | 190.7915 | 248.7800 | 260.5949 | 209.3750 | 217.0923 | 274.2974 | 268.4877 | 273.6359 | 302.2596 |
| $\omega_e$, cm$^{-1}$ | 329.2293 | 627.6227 | 745.7320, | 365.2066 | 474.8314 | 505.7320 | 311.5653 | 342.8103 | 353.3526 |
| | 770.7360 | 902.6134 | 921.8177 | 605.2259 | 763.4506 | 880.7784 | 399.8743 | 833.9337 | 857.0257 |
| | 978.5479 | 1126.9111 | 1644.0542 | 928.3934 | 945.4573 | 1587.9404 | 951.1469 | 962.6116 | 1940.7372 |

Table S7: Cartesian coordinates, rotational constant, zero point energy corrections (ZPE), point group, symmetry number and frequencies ($\omega$) of intermediates complexes in the reaction (4) $RuO_3 + NO_2 \longrightarrow RuO_4 + NO$, reaction path 2

| Properties | MCR (4-P2) | | | TS1 (4-P2) | | | MCP1 (4-P2) | | |
|---|---|---|---|---|---|---|---|---|---|
| ZPE (kJ/mol) | 52.17 | | | 47.51 | | | 47.12 | | |
| Rotational constants (GHZ) | X | Y | Z | X | Y | Z | X | Y | Z |
| | 3.1897811 | 1.9653313 | 1.6809660 | 3.3226968 | 1.7135198 | 1.5555175 | 3.3767902 | 1.5714234 | 1.5201628 |
| Full Point Group | C1 | | | C1 | | | C1 | | |
| Symmetry number | 1 | | | 1 | | | 1 | | |
| COORDINATES | | | | | | | | | |
| O | 0.403810 | 0.000134 | 1.697715 | 0.359829 | 0.512360 | -1.622831 | -0.490762 | -0.920366 | -1.459263 |
| Ru | 0.424408 | -0.000006 | -0.032719 | 0.482317 | 0.040684 | 0.028696 | -0.479087 | -0.072352 | 0.000587 |
| O | 1.097495 | 1.445311 | -0.572515 | 1.628618 | -1.192603 | 0.161477 | -1.661762 | 1.182711 | -0.046640 |
| O | 1.097549 | -1.445361 | -0.572347 | 0.545003 | 1.393756 | 1.031210 | -0.553781 | -0.871274 | 1.485837 |
| O | -1.499206 | -1.033113 | -0.124427 | -2.098768 | 0.676576 | -0.084511 | 2.306062 | -0.625735 | 0.005218 |
| N | -2.211077 | -0.000016 | -0.141739 | -2.243018 | -0.506055 | 0.095244 | 2.306671 | 0.552370 | 0.006089 |
| O | -1.499197 | 1.033077 | -0.124447 | -1.124785 | -1.171051 | 0.273490 | 1.016883 | 1.149275 | 0.006292 |
| | 112.2220 | 167.3905 | 195.6424 | i209.4966 | 83.4630 | 114.7351 | 60.3839 | 101.3216 | 119.4237 |
| | 278.1602 | 310.6052 | 318.8845 | 151.0212 | 215.6688 | 257.5361 | 162.6363 | 255.0706 | 274.9627 |
| $\omega_e$, cm$^{-1}$ | 319.0243 | 331.3265 | 454.3623 | 311.8179 | 351.8017 | 431.7187 | 297.9400 | 351.7756 | 414.6997 |
| | 868.1649 | 900.0689 | 966.3090 | 854.8886 | 867.5116 | 895.9454 | 594.4574 | 808.5766 | 883.5150 |
| | 968.9186 | 1263.5710 | 1266.9160 | 936.7404 | 957.5014 | 1513.1125 | 953.8799 | 954.9926 | 1643.9051 |

Table S8: Cartesian coordinates, rotational constant, zero point energy corrections (ZPE), point group, symmetry number and frequencies ($\omega$) of intermediates complexes in the reaction (4) $RuO_3 + NO_2 \longrightarrow RuO_4 + NO$, reaction path 2

| Properties | RuO$_3$NO$_2$ (4-P2) | | | TS2 (4-P2) | | | MCP (4-P2) | | |
|---|---|---|---|---|---|---|---|---|---|
| ZPE (kJ/mol) | 47.37 | | | 45.61 | | | 46.19 | | |
| | X | Y | Z | X | Y | Z | X | Y | Z |
| Rotational constants (GHz) | 3.6744207 | 1.3620384 | 1.3494825 | 3.7315446 | 1.3562570 | 1.3505825 | 3.8616711 | 1.2054800 | 1.181650 |
| Full Point Group | C1 | | | C1 | | | C1 | | |
| Symmetry number | 1 | | | 1 | | | 1 | | |
| COORDINATES | X | Y | Z | X | Y | Z | X | Y | Z |
| O | 2.621047 | -0.421921 | 0.32149 | 2.622207 | -0.478191 | 0.207961 | 2.855119 | 0.524763 | 0.020547 |
| Ru | -0.565193 | -0.003139 | -0.009998 | -0.554381 | -0.004967 | -0.006087 | -0.568135 | 0.002717 | 0.000238 |
| O | 0.902222 | 0.764956 | -0.817953 | 0.813932 | 0.914497 | -0.738018 | 0.563461 | -1.280915 | -0.008644 |
| O | -1.500292 | 1.222213 | 0.693215 | 0.279298 | -1.060749 | 1.078813 | 0.324837 | 1.446432 | -0.116467 |
| O | 0.198873 | -0.888642 | 1.269538 | -1.336133 | -0.867055 | -1.236135 | -1.625903 | -0.138219 | -1.318681 |
| N | 2.477732 | 0.477794 | -0.351613 | 2.499426 | 0.551009 | -0.234010 | 2.824600 | -0.610979 | -0.020554 |
| O | -1.281306 | -1.077412 | -1.103642 | -1.517205 | 1.036681 | 0.925617 | -1.464294 | -0.032398 | 1.439918 |
| | 52.3420 | 107.7214 | 155.0084 | *i229.8606* | 73.5887 | 117.6528 | 24.2557 | 63.2794 | 85.7601 |
| | 181.5202 | 243.1397 | 282.7730 | 151.7777 | 215.6089 | 258.3733 | 108.9471 | 231.0012 | 281.7769 |
| $\omega_e$, cm$^{-1}$ | 298.2996 | 318.0359 | 329.7275 | 304.1770 | 307.1450 | 336.7765 | 319.1218 | 325.1197 | 335.4788 |
| | 613.5139 | 693.2790 | 872.3389 | 554.7772 | 672.8292 | 857.2355 | 358.5059 | 880.9960 | 932.4629 |
| | 943.0209 | 953.2572 | 1875.9435 | 938.2707 | 954.8374 | 1882.7888 | 954.5541 | 960.2445 | 1860.4660 |

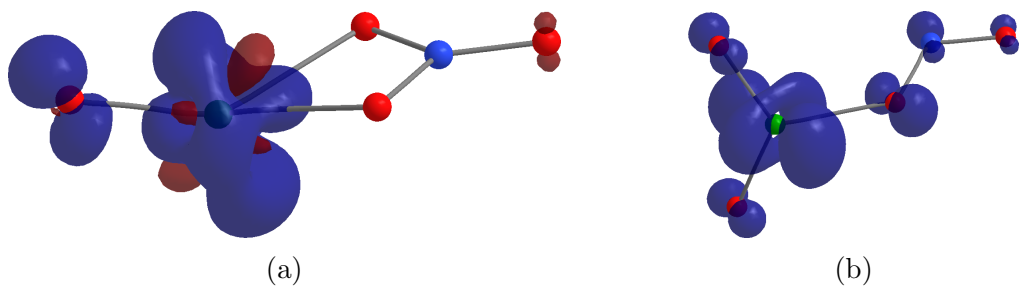

(a) (b)

Figure S2: MCR(3) (a) and MCP(3) (b) with spin density. Red atom = O, blue atom = N, green atom = Ru, blue spin density = $\alpha$, and red spin density = $\beta$.

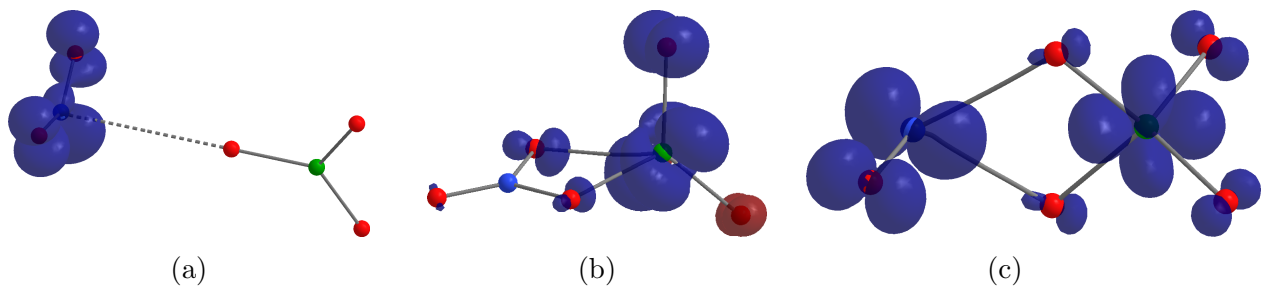

(a) (b) (c)

Figure S3: MCR(4-P1) (a), intermediate species RuO$_2$NO$_3$ (b) and MCR(4-P1) (c) with spin density. See Fig. S2 for color legend.

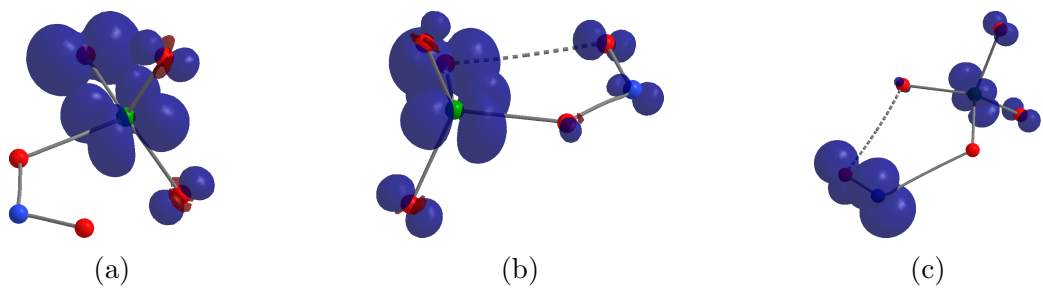

Figure S4: MCR(4-P2) (a), intermediate species $RuO_3NO_2$ (b) and MCR(4-P2) (c) with spin density. See Fig. S2 for color legend.



\end{document}